\documentclass[aps,pra,superscriptaddress,twocolumn,showpacs,amsmath]{revtex4-1}
\usepackage{amssymb,amsmath}
\usepackage{mathrsfs}
\usepackage{graphicx}
\usepackage{float}
\usepackage{txfonts}
\usepackage[position=t,singlelinecheck=off,
caption=false]{subfig}
\usepackage[normalem]{ulem}
\usepackage{color}
\usepackage{bm}
\usepackage{xcolor}
\usepackage{pdfcomment}
\usepackage{graphicx}
\usepackage{dcolumn}
\usepackage{bm}
\usepackage{bbm}
\usepackage{soul}
\setstcolor{red}

\usepackage[T1]{fontenc}
\usepackage[latin9]{inputenc}
\usepackage{setspace}
\usepackage{esint}

\begin{document}

\preprint{APS/123-QED}

\title{Non-Hermitian Floquet topological phases with arbitrarily many real-quasienergy edge states}

\author{Longwen \surname{Zhou}}\email{zhoulw13@u.nus.edu}
\affiliation{Department of Physics, National University of Singapore, Singapore 117551, Republic of Singapore}
\affiliation{Department of Physics, College of Information Science and Engineering, Ocean University of China, Qingdao 266100, China}

\author{Jiangbin \surname{Gong}}\email{phygj@nus.edu.sg}
\affiliation{Department of Physics, National University of Singapore, Singapore 117551, Republic of Singapore}

\date{\today}

\begin{abstract}
Topological states of matter in non-Hermitian systems have
attracted a lot of attention due to their intriguing dynamical and
transport properties. In this study, we propose a periodically driven
non-Hermitian lattice model in one-dimension, which features rich
Floquet topological phases. The topological phase diagram of the model
is derived analytically. Each of its non-Hermitian
Floquet topological phases is characterized by a pair of integer winding numbers,
counting the number of {\it real} $0$- and $\pi$-quasienergy edge states at
the boundaries of the lattice. Non-Hermiticity induced Floquet topological
phases with unlimited winding numbers are found, which allow arbitrarily many {\it real} $0$- and $\pi$-quasienergy edge states to appear in the complex quasienergy bulk gaps in a well-controlled manner.
We further suggest to probe
the topological winding numbers of the system by dynamically imaging
the stroboscopic spin textures of its bulk states.
\end{abstract}

\maketitle

\section{Introduction}\label{sec:Int}

Floquet topological states of matter appear in systems described by
time-periodic Hamiltonians. Being intrinsically out-of-equilibrium,
these states are characterized by topological invariants, bulk-edge
relations and classification schemes that are either similar to or distinct from their
static counterparts~\cite{OkaPRB2009,LindnerNP2011,DahlhausPRB2011,KitagawaPRB2011,HailongPRE2013,LeonPRL2013,CayssolRRL2013,ThakurathiRRB2013,GrushinPRL2014,Wang2014,TitumPRL2015,DLossPRLPRB,KR4,KitagawaPRB2010,JiangPRL2011,KunduPRL2013,Bomantara2016,ZhaoErH2014,ReichlPRA2014,LeonPRB2014,AnomalousESPRX,FulgaPRB2016,ZhouPRB2016,AsbothSTF,NathanNJP2015,ClassificationFTP1,ClassificationFTP2,DerekPRL2012,TongPRB2013,ZhouEPJB2014,XiongPRB2016}. Under properly designed driving fields, Floquet
topological phases with many edge states can be
engineered~\cite{DerekPRL2012,TongPRB2013,ZhouEPJB2014,XiongPRB2016,SeradjehArXiv2017}, resulting in intriguing transport signatures~\cite{YapPRB2017,YapPRB2018}.
Especially, recipes and prototypical models were discovered, which allow the
generation of unlimited number of degenerate $0/\pi$ edge modes~\cite{DerekPRB2014,ZhouPRA2018}, chiral
edge states~\cite{ZhouPRB2018} and linked nodal loops~\cite{LinhuarXiv2018} in well-controlled manners.
Experimentally, Floquet topological states have also been found in cold atom, photonic,
phononic and acoustic systems~\cite{ColdAtomFTP,ColdAtomFTP2,PhotonFTP0,PhotonFTP1,PhotonFTP2,PhononFTP,PhononFTP2}.

In more general situations, the physical setups used to realize Floquet
topological phases are usually subject to gain and loss or environment-induced
dissipations~\cite{NHTPReview}. These processes could be effectively described by introducing
a non-Hermitian part to the Hamiltonian of the Floquet system. The
dynamics of the system, now induced effectively by a time-periodic non-Hermitian
Hamiltonian would then be non-unitary, and the quasienergies
of the corresponding Floquet operator (i.e., time evolution operator
over a complete driving period) could also be complex. How will these
non-Hermitian effects change a Floquet topological state, and whether
non-Hermiticity induced Floquet topological phases with new features
could appear in these systems are largely unexplored~\cite{YuceFloquetPT,KimArXiv2016}.
In some previous studies, periodic drivings have been employed
to stabilize stroboscopic dynamics~\cite{GongPRA2015} and reveal signatures
of exceptional points~(EPs)~\cite{UzdinJPA2011,BerryJPA2011,HeissEP,HuPRB2017,KottosPRA2018,GongPRA2018,ZhouSci2018}.
In another early work, losses have been introduced as dissipative
probes to the topological invariants of a Hermitian lattice model~\cite{RudnerPRL2009}.
This idea motivates further theoretical analysis and experimental
realizations of non-Hermitian quantum walks~\cite{ZeunerPRL2015,MochizukiPRA2016,HuangPRA2016,RakovszkyPRB2017,XiaoNatPhys2017,ChenPRA2018,HarterArXiv2018}, together with the detection
of their topological invariants~\cite{WangArXiv2018}.

In this manuscript, we further reveal the richness of Floquet topological
phases in non-Hermitian systems. We first propose a periodically driven
non-Hermitian lattice model, which could be realizable in various types
of quantum simulators. Under periodic boundary conditions (PBC), the
phase diagram of the model is obtained by deriving the general conditions
for band touching points~(BTPs) to appear in its complex Floquet quasienergy spectrum. Each of the
phases is further shown to be characterized by
a pair of integer topological winding numbers. Under open boundary conditions
(OBC), these winding numbers predict the number of topological edge
states appear at zero and $\pi$-quasienergies in the Floquet spectrum.
Remarkably, our simple model shows that it is possible to have a
non-Hermitian Floquet system with
unlimited winding numbers and arbitrarily many topological edge states with {\it purely real} quasienergies.
Finally, we propose to image the stroboscopic spin textures of our
system, whose patterns have direct connections with the winding numbers
of each of its non-Hermitian Floquet topological phases.

\section{The model}\label{sec:Model}
\begin{figure}
	\centering
	\includegraphics[width=.47\textwidth]{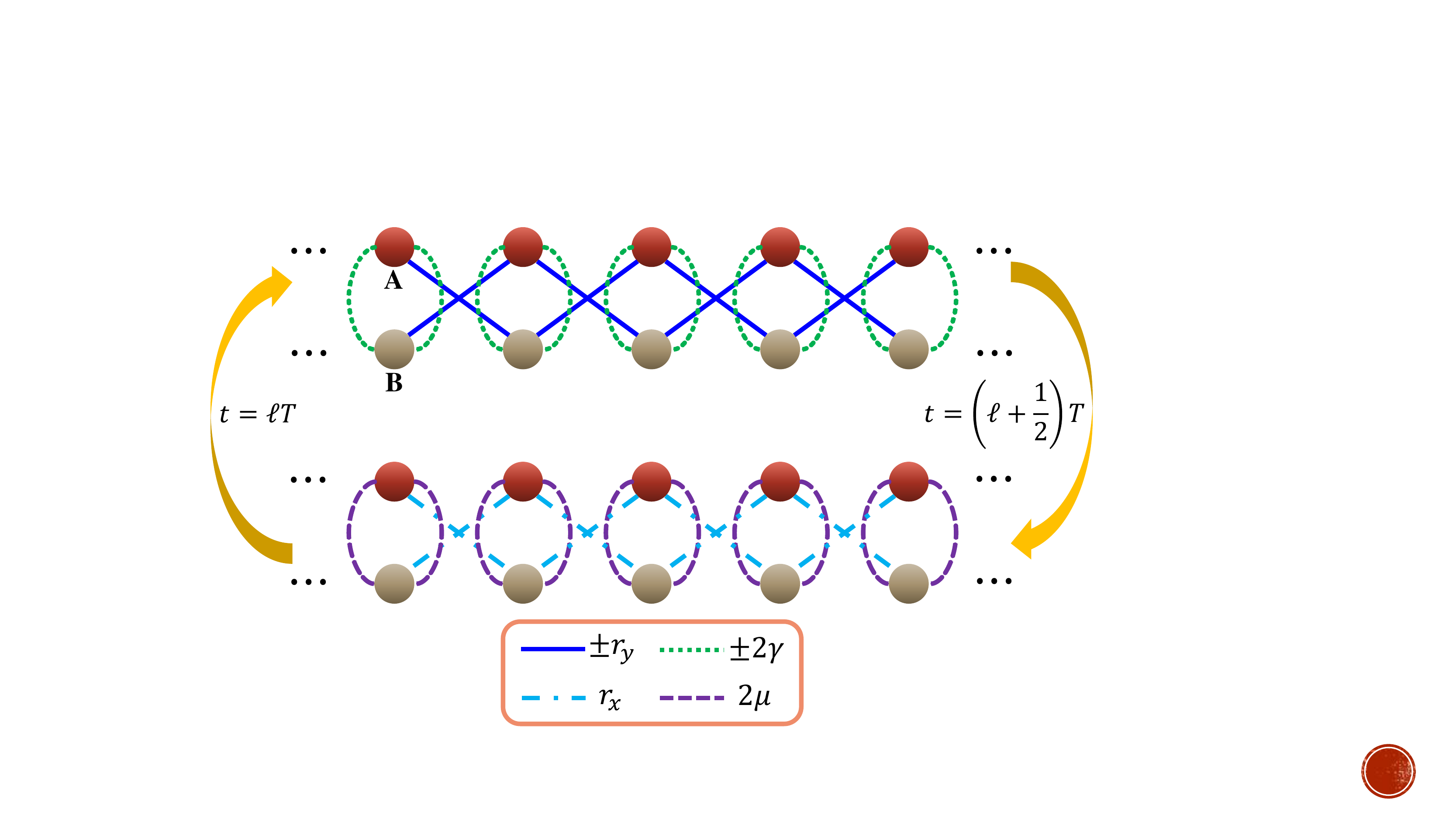}
	\caption{A sketch to the geometry of the periodically driven non-Hermitian
		lattice model Eq. (\ref{eq:H}) over a complete driving period. Each
		of the unit cells contains two sublattices A and B. In the first half
		of the driving period, the system has non-Hermitian intracell couplings
		$\pm2\gamma$ and intercell couplings $\pm r_{y}$ among adjacent
		unit cells. In the second half of the driving period, the system has
		intracell couplings $2\mu$ and intercell couplings $r_{x}$ among
		adjacent unit cells. The same sequence of quenches repeat in each
		driving period.}
	\label{fig:Lattice}
\end{figure}
In this study, we focus on a tight-binding lattice model with a ladder
geometry, which is subjected to piecewise time-periodic quenches.
The geometry of the lattice is illustrated in Fig.~\ref{fig:Lattice}, with two sublattices
A and B in each of its unit cells. The time-dependent Hamiltonian
of the lattice is given by:
\begin{equation}
\hat{H}(t)=\begin{cases}
\hat{H}_1, & t\in\left[\ell T,\ell T+\frac{T}{2}\right).\\
\hat{H}_2, & t\in\left[\ell T+\frac{T}{2},\ell T+T\right).
\end{cases}\label{eq:H}
\end{equation}
Here $\hat{H}_1=\sum_{n}[ir_{y}(|n+1\rangle\langle n|-{\rm h.c.})+2i\gamma|n\rangle\langle n|]\otimes\sigma_{y}$, $\hat{H}_2=\sum_{n}\left[r_{x}(|n\rangle\langle n+1|+{\rm h.c.})+2\mu|n\rangle\langle n|\right]\otimes\sigma_{x}$, $\ell\in\mathbb{Z}$, $T$ is the driving period, $n\in\mathbb{Z}$
is the unit cell index and the lattice constant has been set to $1$.
The system parameters $r_{x},r_{y},\mu$ and $\gamma$ all take real
values. The Pauli matrices $\sigma_{x},\sigma_{y}$ act on sublattice
degrees of freedom A and B.

In the first half of a driving period, the lattice Hamiltonian contains
intercell hoppings $\pm r_{y}$ between A, B sublattices in nearest
neighbor unit cells, and asymmetric couplings $\pm2\gamma$ between
sublattices A,B in the same unit cell, which introduce non-Hermitian
effects. In experimental setups like coupled-resonator waveguides,
such asymmetric couplings may be realized by the asymmetric scattering
between a clockwise and a counterclockwise propagating mode within each
resonator~\cite{MalzardPRL2015}.
These two resonator modes correspond to the A, B sublattices. Their internal coupling can be made asymmetric by opening the system, resulting in asymmetric internal scattering and losses within the cavity. In alternative experimental setups like cold atoms in optical lattices, the A, B sublattices are replaced by spins of cold atoms. Under a simple rotation $\sigma_y\rightarrow\sigma_z$ yielding a mathematically equivalent model, the non-Hermitian term becomes $i\gamma\sigma_z$, which is associated with particle gain/loss for spin up/down atoms (assuming $\gamma>0$). Experimentally, it may be realized by introducing an atom loss of strength $-2i\gamma$ for spin down atoms only, which could be generated using a resonant optical beam to kick the atoms in the spin-down state out of a trap. One may also consider applying a radio frequency pulse to excite atoms in the spin-down state to another irrelevant state, leading to an effective decay when atoms in that state experience a loss by applying an antitrap~\cite{XuPRL2017}.

In the second half of
a driving period, the lattice Hamiltonian is Hermitian, with hoppings
$2\mu$ and $r_{x}$ between sublattices A and B in the same
unit cell and among nearest neighbor unit cells, respectively.
This relatively simple lattice model should also be realizable in both coupled resonator and cold atom systems.
Putting
together, the Floquet operator describing the evolution of the system
over a complete driving period is given by
\begin{alignat}{1}
\hat{U}= & \,e^{-i\sum_{n}\left[\frac{r_{x}}{2}(|n\rangle\langle n+1|+{\rm h.c.})+\mu|n\rangle\langle n|\right]\otimes\sigma_{x}}\nonumber \\
\times & \,e^{-i\sum_{n}\left[\frac{r_{y}}{2i}(|n\rangle\langle n+1|-{\rm h.c.})+i\gamma|n\rangle\langle n|\right]\otimes\sigma_{y}},\label{eq:U}
\end{alignat}
where we have set $\hbar=T=1$ and chosen $\hbar/T$ to be the unit
of energy. Under PBC, the Floquet operator $\hat{U}$ can also be expressed
in momentum representation as $\hat{U}=\sum_{k}U(k)|k\rangle\langle k|$,
where
\begin{alignat}{1}
U(k)= & \,e^{-ih_{x}(k)\sigma_{x}}e^{-i[h_{y}(k)+i\gamma]\sigma_{y}},\label{eq:Uk}\\
h_{x}(k)= & \,\mu+r_{x}\cos(k),\label{eq:hxk}\\
h_{y}(k)= & \,r_{y}\sin(k),\label{eq:hyk}
\end{alignat}
and $k\in[-\pi,\pi)$ is the quasimomentum. Without periodic drivings,
topological phases in similar non-Hermitian lattice models have been
explored in several studies~\cite{LeePRL2016,WeimannNatMat2017,LeykamPRL2017,ZhouArXiv2017,ShenPRL2018,YinPRA2018,RyuPRB2017,YucePRA2018,YaoArXiv2018,LieuPRB2018}. In the following, we will show that the
periodic quenches considered in this manuscript make the system much
richer in realizing non-Hermitian topological states with real quasienergies, with the possibility
of reaching phases with arbitrarily large topological invariants and
induced solely by non-Hermitian effects.

\section{Bulk properties}\label{sec:Bulk}
To reveal the richness of Floquet topological phases in the periodically
driven non-Hermitian lattice (PDNHL) model introduced in Eq.~(\ref{eq:H}),
we first analyze the bulk quasienergy spectrum and eigenstates of
its Floquet operator $U(k)$ in the following subsections. By investigating
the gapless condition of the quasienergy spectrum, we obtain the trajectories
of Floquet BTPs in the parameter space, which form the boundaries of
different non-Hermitian Floquet topological phases. We further introduce
a pair of integer winding numbers, which could uniquely characterize each
of the Floquet topological phases in the phase diagram.

\subsection{Floquet spectrum and band-touching points}\label{sec:FEP}

The Floquet spectrum of $U(k)$ as defined in Eq.~(\ref{eq:Uk}) has
two quasienergy bands $\pm E(k)$~\cite{cnote1}, with $E(k)$ being an eigenvalue
of the effective Floquet Hamiltonian $H(k)=-i\ln U(k)$. Since $H(k)$
is non-Hermitian, we in general have $E(k)\neq E^{*}(k)$. But similar
to Hermitian Floquet systems, the real part of $E(k)$ is a phase
factor, which is defined modulus $2\pi$ and take values in the first
quasienergy Brillouin zone $(-\pi,\pi]$. Therefore the Floquet spectrum
of $U(k)$ could have two gaps around $E(k)=0$ and $E(k)=\pi$ in
the complex quasienergy plane.

Combining the two parts of quenches in Eq.~(\ref{eq:Uk}),
$U(k)$ can be written as $U(k)=e^{-iE(k)\hat{\mathbf n}\cdot{\boldsymbol \sigma}}$,
where $\hat{\mathbf n}$ is a unit vector and ${\boldsymbol \sigma}=(\sigma_x,\sigma_y,\sigma_z)$.
Then using the formula $e^{i\theta\hat{\mathbf n}\cdot{\boldsymbol \sigma}}=\cos(\theta)+i\sin(\theta)\hat{\mathbf n}\cdot{\boldsymbol \sigma}$ to Eq.~(\ref{eq:Uk}),
it is not hard to see that $E(k)$ satisfies
the equation:
\begin{equation}
\cos[E(k)]=\cos[h_{x}(k)]\cos[h_{y}(k)+i\gamma].\label{eq:cosEk}
\end{equation}
Then if the quasienergy gap closes at $E(k)=0$ or $E(k)=\pi$, we
will have $\cos[E(k)]=1$ or $\cos[E(k)]=-1$, respectively. Plugging
Eqs.~(\ref{eq:hxk}) and (\ref{eq:hyk}) into the right hand side
of Eq.~(\ref{eq:cosEk}), we find the gapless conditions to be $\mu+r_{x}\cos(k)=m\pi\pm\arccos\left[\frac{1}{\cosh(\gamma)}\right]$
and $r_{y}\sin(k)=n\pi$, where $m,n$ are integers of the same parity
(opposite parities) if the gap closes at $E(k)=0$ {[}$E(k)=\pi${]}.
Combining these gapless conditions together
with the help of the identity $\cos^2(k)+\sin^2(k)=1$,
we find the equation for trajectories
of Floquet BTPs (i.e., BTPs of the complex quasienergy
spectrum) in the parameter space as:
\begin{equation}
\frac{1}{r_{x}^{2}}\left\{ m\pi\pm\arccos\left[\frac{1}{\cosh(\gamma)}\right]-\mu\right\} ^{2}+\frac{n^{2}\pi^{2}}{r_{y}^{2}}=1,\label{eq:FEP}
\end{equation}
where $m,n\in\mathbb{Z}$. These trajectories form boundaries separating
different non-Hermitian Floquet topological phases, as will be discussed
in the following subsections.

\subsection{Phase boundaries}\label{sec:PhsBound}
Before presenting explicit examples of the phase diagram at finite
values of the non-Hermitian coupling strength $\gamma$, let's first
discuss two limiting cases. In the Hermitian limit ($\gamma=0$),
Eq.~(\ref{eq:FEP}) reduces to $\frac{(m\pi-\mu)^{2}}{r_{x}^{2}}+\frac{n^{2}\pi^{2}}{r_{y}^{2}}=1$.
The topological phases in the corresponding Hermitian Floquet system
have been explored in Ref.~\cite{ZhouPRA2018}, with very rich Floquet topological
states identified theoretically in the context of a spin-$1/2$ kicked
rotor. In the opposite limit ($\gamma\rightarrow\infty$), Eq.~(\ref{eq:FEP})
is simplified to $\frac{(m\pi\pm\frac{\pi}{2}-\mu)^{2}}{r_{x}^{2}}+\frac{n^{2}\pi^{2}}{r_{y}^{2}}=1$.
In this case, Floquet BTPs appearing at zero
and $\pi$ quasienergies tend to coincide with each other in the parameter
space, resulting in simpler phase boundary structures. For
completeness, we present examples of phase boundary diagrams near
these two limits in the Appendix~\ref{app:PhsBound}.

\begin{figure}
	\centering
	\includegraphics[width=.46\textwidth]{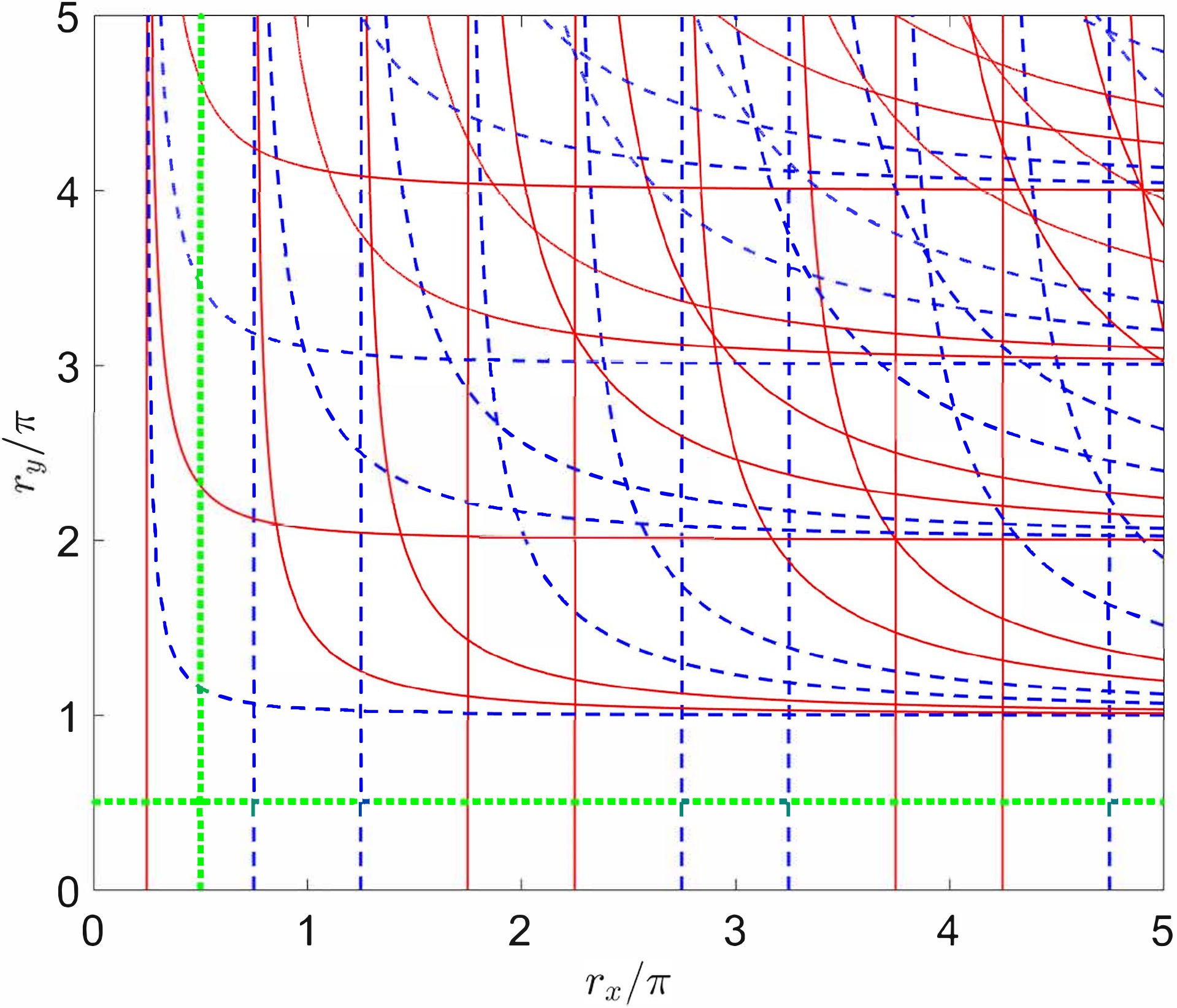}
	\caption{Phase boundary diagram of the PDNHL model in the parameter space $(r_{x},r_{y})\in(0,5\pi]\times(0,5\pi]$,
		with other system parameters fixed at $\mu=0$ and $\gamma={\rm arccosh}(\sqrt{2})$.
		Trajectories of Floquet BTPs corresponding to spectrum gap closings
		at quasienergy zero ($\pi$) are denoted by red solid (blue dashed)
		lines, separating potentially different non-Hermitian Floquet topological phases. Green dotted lines represent parameter sets in which the
		topological properties of the system are further explored in later sections.}
	\label{fig:PhsBound}
\end{figure}

For a general non-Hermitian coupling $\gamma\neq0$, the phase boundaries (gap-closing lines) separating potentially differet topological phases
are splitted and deformed from the Hermitian limit by an amount $\sim\pm\arccos\left[\frac{1}{\cosh(\gamma)}\right]$.
Note that these changes cannot be eliminated by tuning
the strength of intracell hopping $\mu$, which indicates their genuine
non-Hermitian origin. In Fig.~\ref{fig:PhsBound}, we present an example of the phase
boundary diagram (formed by gap-closing lines in the $r_x$-$r_y$ parameter space) with $\mu=0$ at a fixed $\gamma={\rm arccosh}(\sqrt{2})$.
Floquet BTPs forming the red solid~(blue dashed) lines are related
to spectrum gap closings at quasienergy zero ($\pi$), separating potentially different non-Hermitian Floquet topological phases. We further
notice that for any given integers $m$ and $n$, the Floquet BTPs
related to the two branches $\pm\arccos\left[\frac{1}{\cosh(\gamma)}\right]$ of phase boundaries
are always due to gap closings at the same quasienergy (either
zero or $\pi$). In between these two branches, new Floquet topological
phases that are absent in the Hermitian limit may emerge. In the next
subsection, we will give a unique characterization of the phase formed
in each closed patch on the phase boundary diagram through a pair
of topological invariants.

\subsection{Topological winding number and phase diagram}\label{sec:WN}
The symmetry classification of Hermitian Floquet topological states
has been established in several studies~\cite{ClassificationFTP1,ClassificationFTP2}. In one dimension, all Floquet
topological states in Hermitian systems are symmetry protected. One
important class of these topological phases is that protected by a
chiral symmetry. Each one-dimensional Floquet topological phase with
chiral symmetry is characterized by a pair of integer winding numbers,
defined in two symmetric time frames of the system's Floquet operator~\cite{AsbothSTF}.
For non-Hermitian Floquet systems, however, a general symmetry classification
scheme has not yet been formulated. Our study here provides useful resources for the
establishment of such a framework.

The Floquet operator $U(k)$ of our PDNHL model in Eq.~(\ref{eq:Uk}) has chiral symmetry in two symmetric
time frames. These frames are obtained by shifting the starting time
of the evolution forward or backward over half of the driving period.
The resulting Floquet operators in these symmetric time frames are
given by:
\begin{alignat}{1}
U_{1}(k)= &\, e^{-i\frac{h_{x}(k)}{2}\sigma_{x}}e^{-i[h_{y}(k)+i\gamma]\sigma_{y}}e^{-i\frac{h_{x}(k)}{2}\sigma_{x}},\label{eq:U1k}\\
U_{2}(k)= &\, e^{-i\frac{h_{y}(k)+i\gamma}{2}\sigma_{y}}e^{-ih_{x}(k)\sigma_{x}}e^{-i\frac{h_{y}(k)+i\gamma}{2}\sigma_{y}}.\label{eq:U2k}
\end{alignat}
Note that $U_{1}(k)$ and $U_{2}(k)$ are differ from $U(k)$ by similarity
transformations, and therefore sharing with it the same complex quasienergy
spectrum. However, these Floquet operators in different time frames
are not unitarily equivalent due to the non-Hermiticity of our system.
The chiral symmetry of $U_{1}(k)$ and $U_{2}(k)$ is described
by a unitary transformation $\Gamma=\Gamma^{\dagger}=\Gamma^{-1}$, under which
\begin{equation}
\Gamma U_{\alpha}(k)\Gamma=U_{\alpha}^{-1}(k),\qquad\alpha=1,2.
\end{equation}
It is not hard to see that $\Gamma=\sigma_{z}$ satisfies this condition,
and therefore describes the chiral symmetry of our PDNHL model.

With the chiral symmetry, we can introduce winding numbers for $U_{1}(k)$
and $U_{2}(k)$ following the same routine as in Hermitian Floquet
systems~\cite{AsbothSTF}. Using the Euler formula and Eq.~(\ref{eq:cosEk}),
we can rewrite $U_{1}(k)$ and $U_{2}(k)$ as
\begin{equation}
U_{\alpha}(k)=\cos[E(k)]-i[n_{\alpha x}(k)\sigma_{x}+n_{\alpha y}(k)\sigma_{y}],\label{eq:U12k}
\end{equation}
where $\alpha=1,2$, and the components $n_{\alpha x}(k)$ and $n_{\alpha y}(k)$
are given by:
\begin{alignat}{1}
n_{1x}(k)= & \sin[h_{x}(k)]\cos[h_{y}(k)+i\gamma],\\
n_{1y}(k)= & \sin[h_{y}(k)+i\gamma],\\
n_{2x}(k)= & \sin[h_{x}(k)],\\
n_{2y}(k)= & \cos[h_{x}(k)]\sin[h_{y}(k)+i\gamma].
\end{alignat}
Notably, due to the chiral symmetry, both the real and imaginary parts of vectors ${\bf n}_{1}(k)=[n_{1x}(k),n_{1y}(k)]$
and ${\bf n}_{2}(k)=[n_{2x}(k),n_{2y}(k)]$ are constrained to move
in a two dimensional plane under the change of $k$. Therefore they
have well defined winding numbers around the origin of the plane,
given by
\begin{equation}
W_{\alpha}=\int_{-\pi}^{\pi}\frac{dk}{2\pi}\frac{n_{\alpha x}(k)\partial_{k}n_{\alpha y}(k)-n_{\alpha y}(k)\partial_{k}n_{\alpha x}(k)}{n_{\alpha x}^{2}(k)+n_{\alpha y}^{2}(k)}\label{eq:W12}
\end{equation}
for $\alpha=1,2$. Even though the integrand of these winding numbers
have non-vanishing imaginary parts, it has been shown that these imaginary
parts in general have no windings under the integral of quasimomentum
$k$ over the first Brillouin zone~\cite{YinPRA2018}. Therefore both
$W_{1}$ and $W_{2}$ are real and take integer values. Combining
these winding numbers allows us to introduce another pair of winding
numbers $W_{0}$ and $W_{\pi}$, defined as
\begin{equation}
W_{0}=\frac{W_{1}+W_{2}}{2},\qquad W_{\pi}=\frac{W_{1}-W_{2}}{2}.\label{eq:W0P}
\end{equation}
These winding numbers will be used to characterize the non-Hermitian
Floquet topological phases of our system.

\begin{figure}
	\centering
	\includegraphics[width=.46\textwidth]{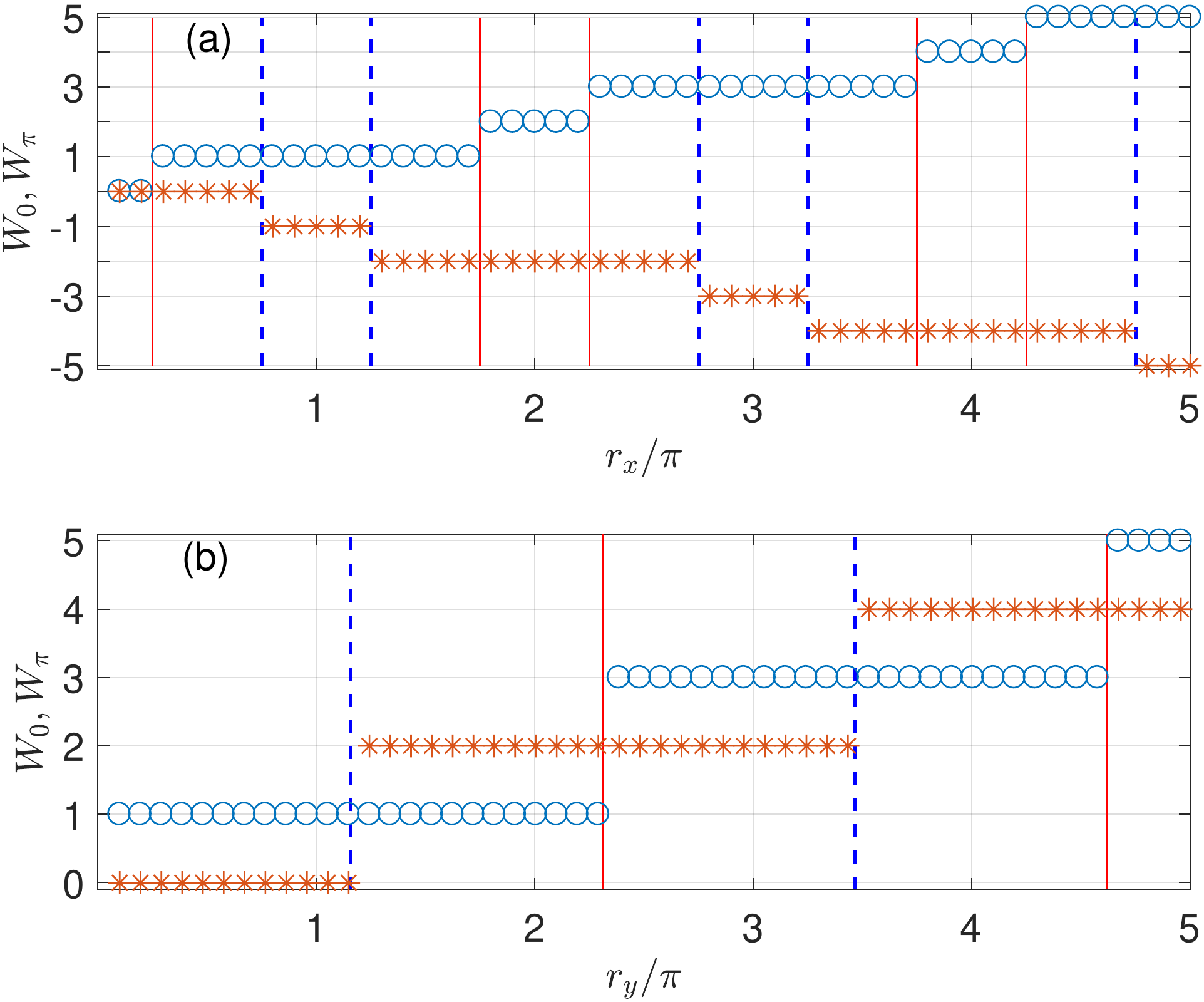}
	\caption{Winding numbers $W_{0}$ (circles) and $W_{\pi}$ (stars) versus system
		parameters $r_{x}$ {[}in panel (a){]} and $r_{y}$ {[}in panel (b){]},
		calculated along the two green dotted lines in the phase boundary
		Fig.~\ref{fig:PhsBound}. Red solid (blue dashed) lines denote phase boundaries mediated
		by spectrum gap closings at quasienergy zero ($\pi$), which are obtained
		by solving Eq. (\ref{eq:FEP}) under the condition $n=0$ ($m=0$).}
	\label{fig:WN}
\end{figure}

As an example, we present in Fig.~\ref{fig:WN} the calculations of $W_{0}$~(blue circles) and $W_{\pi}$~(red stars) along two trajectories (dotted lines) in the phase boundary diagram Fig.~\ref{fig:PhsBound}. We see that in both cases, the winding numbers only take integer values, as suggested by our
theory. Furthermore, the value of $W_{0}$ ($W_{\pi}$) has a quantized
jump every time when a trajectory of Floquet BTPs related to gap closings
at quasienergy zero ($\pi$) is crossed in the parameter space. These
observations indicate that the trajectories of Floquet BTPs described
by Eq.~(\ref{eq:FEP}) are indeed boundaries of different non-Hermitian
Floquet topological phases characterized by different values of winding
numbers $(W_{0},W_{\pi})$. Comparing Fig.~\ref{fig:WN}(a) with the phase diagram
Fig.~1 reported in Ref.~\cite{ZhouPRA2018}, which studied the Floquet topological
phases of our model in its Hermitian limit, we also find new topological
phases in the range of system parameters $n\pi-\arccos\left[\frac{1}{\cosh(\gamma)}\right]<r_{x}<n\pi+\arccos\left[\frac{1}{\cosh(\gamma)}\right]$
for all $n\in\mathbb{Z}$, which are solely induced by non-Hermitian
effects ($\gamma\neq0$). Furthermore, these non-Hermiticity induced Floquet topological
phases could possess arbitrarily large winding numbers. For example,
in the range of parameters $r_{x}\in(n\pi-\arccos\left[\frac{1}{\cosh(\gamma)}\right],n\pi+\arccos\left[\frac{1}{\cosh(\gamma)}\right])$,
$r_{y}\in(0,\pi)$ and for any $\gamma\in(0,\infty)$, we have $(W_{0},W_{\pi})=(n,-n)$
for any $n\in\mathbb{Z}^{+}$.

\begin{figure}
	\centering
	\includegraphics[width=.46\textwidth]{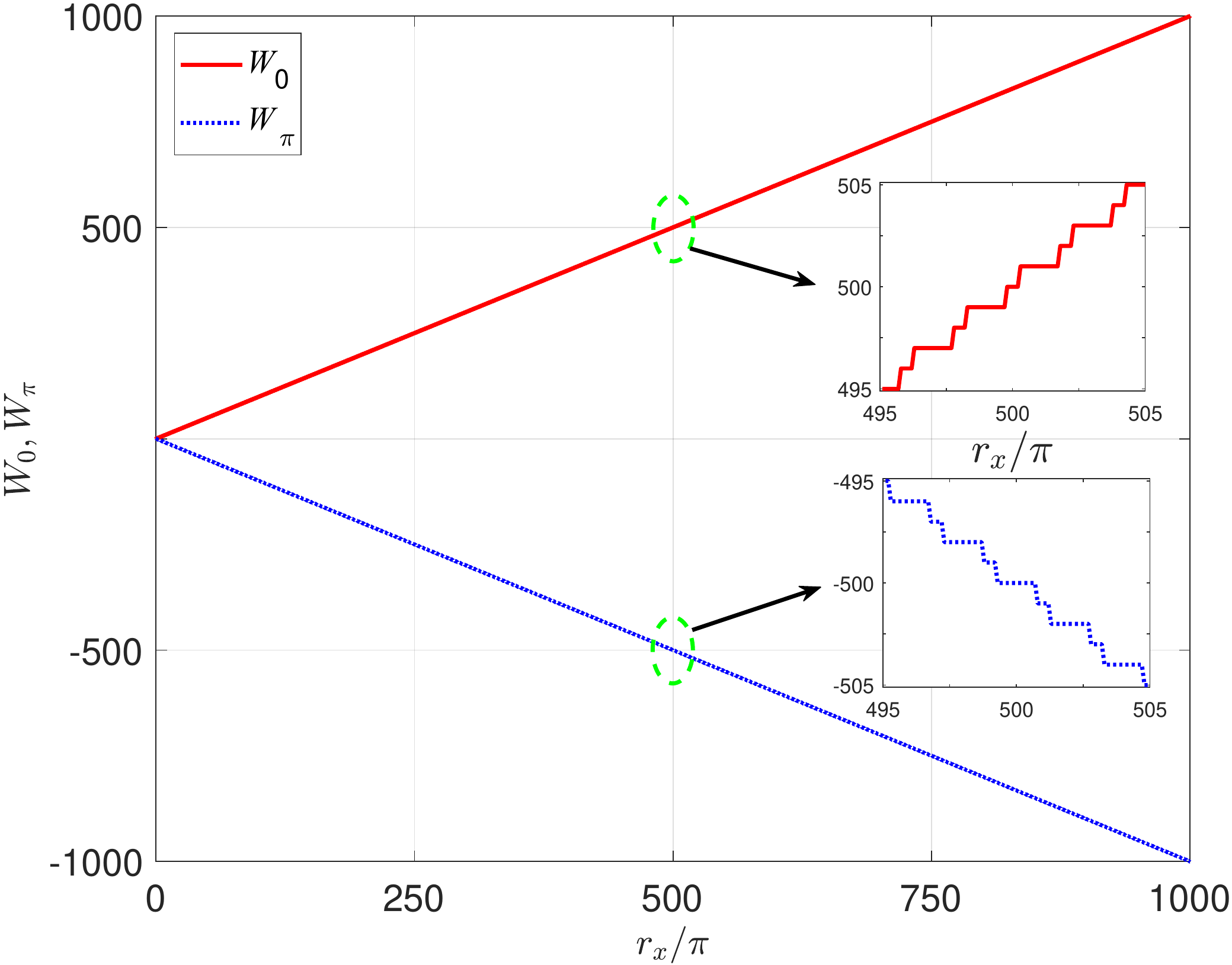}
	\caption{The linear growth of winding numbers $W_{0}$ (red solid line) and
		$W_{\pi}$ (blue dotted line) versus system parameters $r_{x}$, with
		other system parameters fixed at $\mu=0$, $r_{y}=\frac{\pi}{2}$
		and $\gamma={\rm arccosh}(\sqrt{2})$. Green dashed circles highlight
		the regions in which zoomed in results of $W_{0}$ and $W_{\pi}$
		are shown in internal panels.}
	\label{fig:WNgrowth}
\end{figure}

To demonstrate the later point, we present in Fig.~\ref{fig:WNgrowth} the calculation of $(W_{0},W_{\pi})$
as defined in Eq.~(\ref{eq:W0P}) versus the hopping
amplitude $r_{x}$, with other system parameters fixed at $\mu=0$,
$r_{y}=\frac{\pi}{2}$ and $\gamma={\rm arccosh}(\sqrt{2})$. The linear
and unbounded growth of $W_{0}$ and $W_{\pi}$ are clearly seen with
the increase of $r_{x}$. Note that the non-Hermitian topological
phases appearing in the region $r_{x}\in(n\pi-\arccos\left[\frac{1}{\cosh(\gamma)}\right],n\pi+\arccos\left[\frac{1}{\cosh(\gamma)}\right])$
for any $n\in\mathbb{Z}$ are absent in the Hermitian limit ($\gamma=0$)
of the system.

\begin{figure}
	\centering
	\includegraphics[width=.47\textwidth]{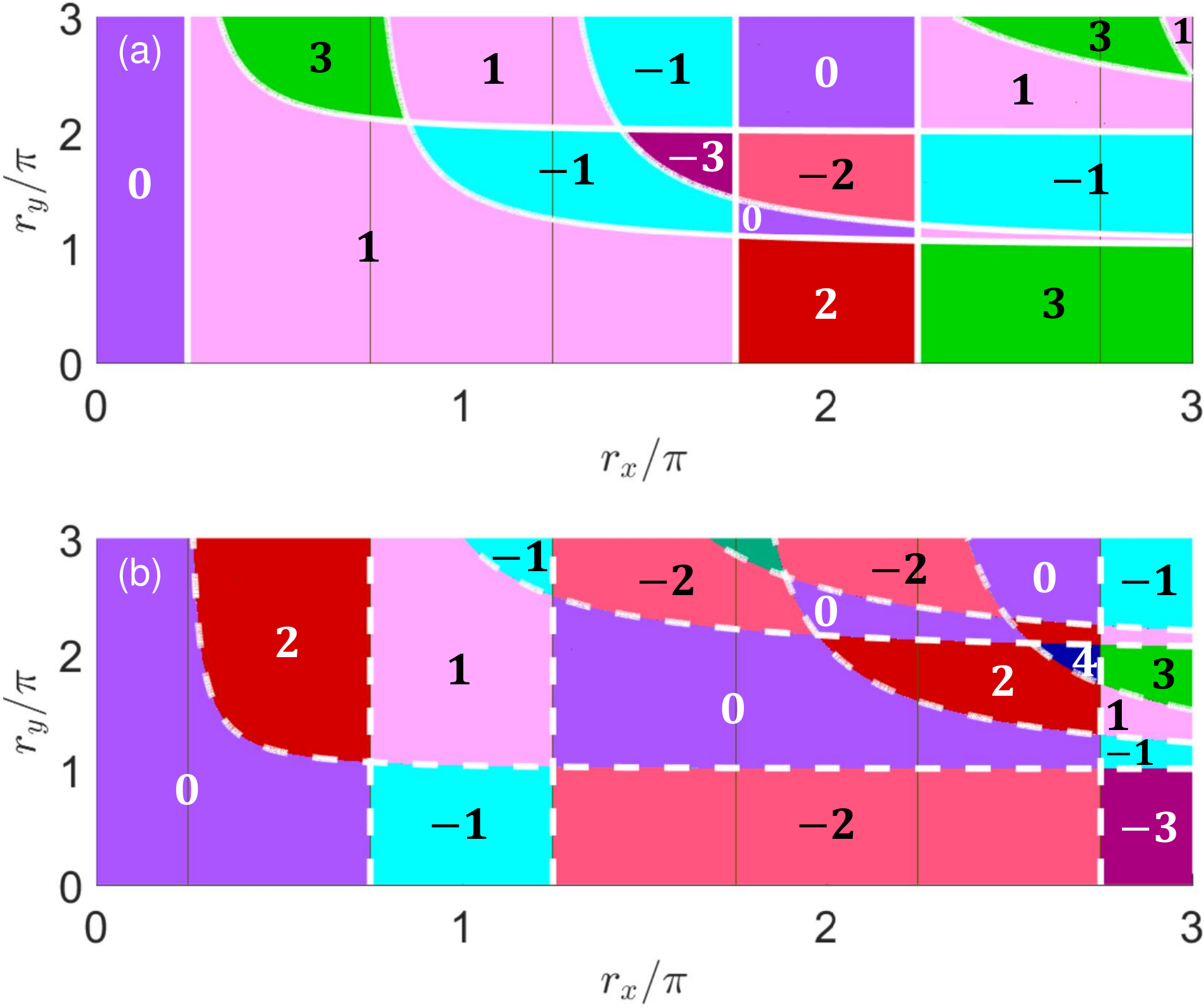}
	\caption{Winding numbers $W_{0}$ {[}panel (a){]} and $W_{\pi}$ {[}panel (b){]}
		in the range of system parameters $(r_{x},r_{y})\in(0,3\pi)\times(0,3\pi)$,
		with other parameters fixed at $\mu=0$ and $\gamma={\rm arccosh}(\sqrt{2})$.
		In panel (a)\textbackslash{}(b), regions with the same color share
		the same values of $W_{0}$\textbackslash{}$W_{\pi}$, as also denoted
		by the integers in the panels. The white solid\textbackslash{}dashed
		lines in panel (a)\textbackslash{}(b) are trajectories of Floquet
		BTPs obtained from Eq. (\ref{eq:FEP}) and related to gap closings
		at quasienergy $0$\textbackslash{}$\pi$. They match precisely with
		the topological phase boundaries at which $W_{0}$ or $W_{\pi}$ gets
		quantized jumps.}
	\label{fig:PhsDiag}
\end{figure}

To give a more global view of the topological phases that can appear in our system, we show in Fig.~\ref{fig:PhsDiag} the results of $W_{0}$
and $W_{\pi}$ in the range of system parameters $(r_{x},r_{y})\in[0,3\pi]\times[0,3\pi]$,
with $\mu=0$ and the non-Hermitian coupling strength $\gamma={\rm arccosh}(\sqrt{2})$.
In panels \ref{fig:PhsDiag}(a) and \ref{fig:PhsDiag}(b), each color corresponds to a range of system parameters
$(r_{x},r_{y})$ in which $W_{0}$ or $W_{\pi}$ takes the same value,
as also specified in the figure. The trajectories of Floquet BTPs corresponding
to gap closings at quasienergies zero (solid lines) and $\pi$ (dashed
lines) are also denoted in the figure. They are found to match precisely
the boundaries across which the values of $W_{0}$ or $W_{\pi}$ take
quantized changes. Therefore we conclude that the trajectories of
Floquet BTPs in the parameter space, as predicted by Eq. (\ref{eq:FEP}),
are indeed boundaries of different non-Hermitian Floquet topological
phases characterized by different values of winding numbers $(W_{0},W_{\pi})$.

The next question to ask concerns the physical implication of these
winding numbers. In the following two sections, we try to address
this issue from two complementary perspectives: the edge states and
bulk dynamics.

\section{Edge states}\label{sec:EdgeStat}
In one dimensional chiral symmetric Hermitian Floquet systems,
the winding numbers $W_{0}$ and $W_{\pi}$ have a direct connection
with the number of degenerate edge state pairs $n_{0}$ and $n_{\pi}$
at quasienergies zero and $\pi$ \cite{AsbothSTF}, i.e.,
\begin{equation}
n_{0}=|W_{0}|,\qquad n_{\pi}=|W_{\pi}|.\label{eq:BEC}
\end{equation}
This relation belongs to the family of bulk-edge correspondence in
topological insulators. Experimentally, it also provides a useful route
to detect topological invariants of bulk materials through quantized
transport along their boundaries, and to distinguish topologically
distinct states of matter by imagining bound states appearing at their
interfaces.

In non-Hermitian systems, however, the relation between bulk invariants
and edge states becomes a subtle issue due to the existence of EPs,
which could make the Hamiltonian matrix of the system defective (algebraic
multiplicity $\neq$ geometric multiplicity), with its spectrum extremely
sensitive to boundary conditions~\cite{NHTPReview}. Cases regrading the breakdown of
bulk-edge correspondence and its reparation in non-Hermitian systems
have been explored in several studies~\cite{LeePRL2016,ShenPRL2018,YinPRA2018,YaoArXiv2018,LieuPRB2018,HuPRB2011,EsakiPRB2011,XiongJPC2018,AlvarezPRB2018,GongArXiv2018,BergholtzArXiv2018}, with still unsettled debates
over the literature. The PDNHL
model we introduced in Eq.~(\ref{eq:H}) provides an example, in which
the bulk-edge relation (\ref{eq:BEC}) for Hermitian Floquet systems
still holds, with all topological edge states taking real quasienergies

\begin{figure}
	\centering
	\includegraphics[width=.47\textwidth]{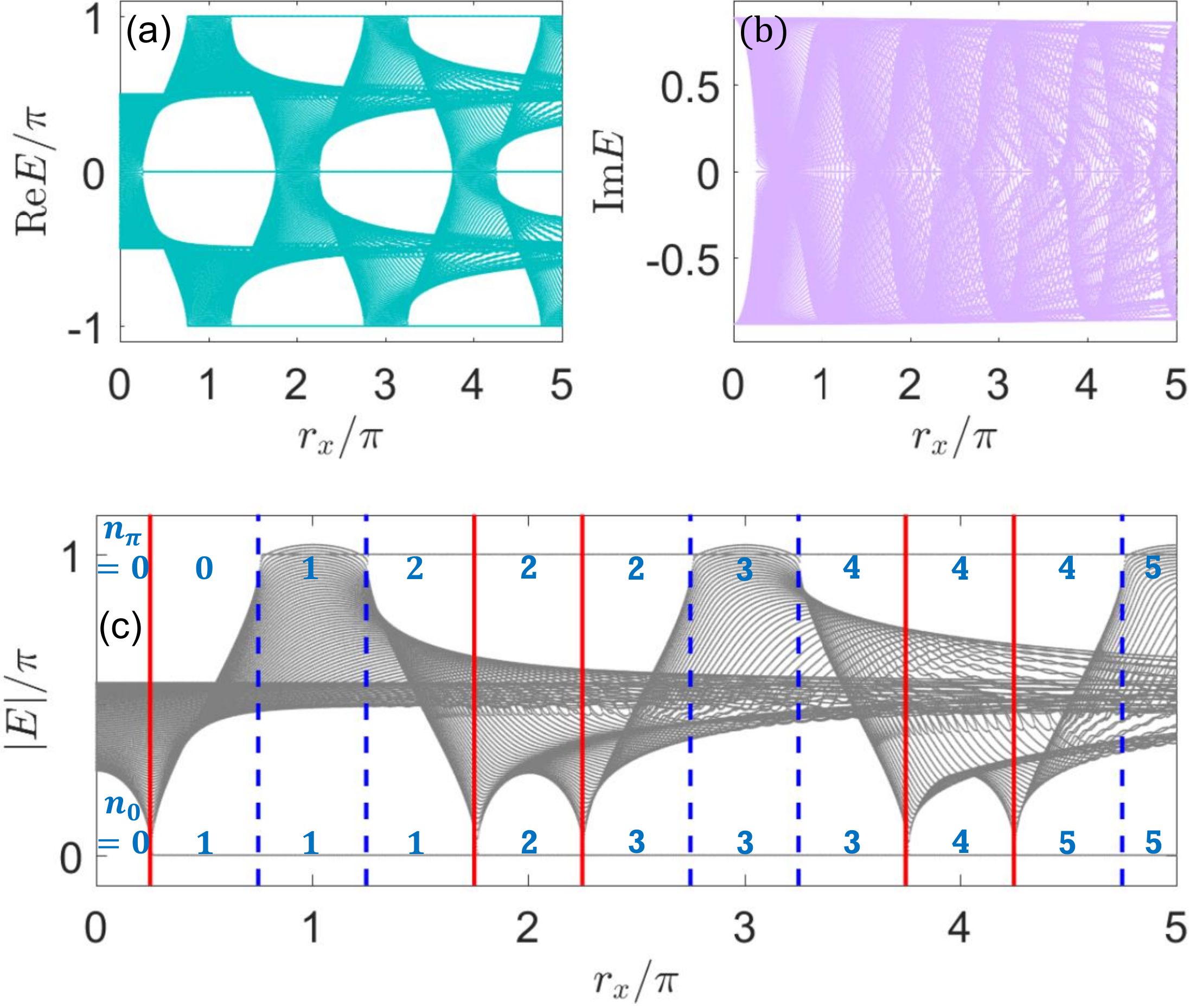}
	\caption{The Floquet spectrum of $\hat{U}$ versus hopping amplitude $r_{x}$
		under OBC. The lattice has $N=150$ unit cells.
		Other system parameters are fixed at $\mu=0$, $r_{y}=\frac{\pi}{2}$
		and $\gamma={\rm arccosh}(\sqrt{2})$. Panels (a) and (b) show the
		real and imaginary parts of the quasienergy $E$. Panel (c) shows
		the absolute values of quasienergy $E$. Red solid (blue dashed) lines
		represent phase boundaries at which the spectrum gap close at quasienergy
		zero ($\pi$). They are obtained from Eq. (\ref{eq:FEP}) for Floquet
		BTPs. The integers in light blue in panel (c) denote the number
		of degenerate edge state pairs at quasienergies zero ($n_{0}$) and
		$\pi$ ($n_{\pi}$). Their values are equal to the absolute values
		of winding numbers $W_{0}$ and $W_{\pi}$ as shown in Fig.~\ref{fig:WN}(a),
		respectively.}
	\label{fig:SpectrumOBC}
\end{figure}

To show this, we studied the quasienergy spectrum and edge states
of Floquet operator $\hat{U}$ as defined in Eq.~(\ref{eq:U}). Numerically,
they are obtained by solving the eigenvalue equation $\hat{U}|\psi\rangle=e^{-iE}|\psi\rangle$
under OBC. As an example, we presented in Fig.~\ref{fig:SpectrumOBC}
the Floquet spectrum $E$ versus hopping amplitude $r_{x}$ at fixed
values of hopping amplitude $r_{y}=\frac{\pi}{2}$ and non-Hermitian
coupling strength $\gamma={\rm arccosh}(\sqrt{2})$ in a lattice of
$N=150$ unit cells. In Fig.~\ref{fig:SpectrumOBC}(c), the red solid and blue dashed lines
correspond to phase boundaries obtained from the Eq. (\ref{eq:FEP})
for Floquet BTPs. We see that the number of edge state pairs $(n_{0},n_{\pi})$,
as denoted in Fig.~\ref{fig:SpectrumOBC}(c), changes across each of the phase boundaries.
Furthermore, refer to Fig.~\ref{fig:PhsDiag}, we find that $(n_{0},n_{\pi})$ matches
exactly the absolute values of winding number $(|W_{0}|,|W_{\pi}|)$
in each of the corresponding non-Hermitian Floquet topological phases.
Therefore the relation (\ref{eq:BEC}) is verified for the example
considered here. With the increase of $r_x$, the bulk-edge relation (\ref{eq:BEC}) and unlimited
winding numbers as shown in Fig.~\ref{fig:WNgrowth} then allow arbitrarily many zero and $\pi$ topological
edge states to appear with purely real quasienergies. This gives the first example of generating many topological
edge states in non-Hermitian systems following the Floquet engineering approach.

In Appendix~\ref{app:EdgeStat}, we present an example of the quasienergy
spectrum $E$ versus $r_{y}$, for which the bulk-edge relation (\ref{eq:BEC})
again survives the test.
In more general situations (e.g., along the direction $r_x=r_y$ in Fig.~\ref{fig:PhsBound}),
the verification of Eq.~(\ref{eq:BEC}) in our PDNHL
is more demanding due to the complicated configuration of BTPs
in the phase diagram, which can result in many topological phase transitions and
fluctuations of the edge state numbers in a relatively small parameter window.
A more systematic understanding of these situations may require statistical-type treatments,
and will be left for future studies.

\begin{figure}
	\centering
	\includegraphics[width=.47\textwidth]{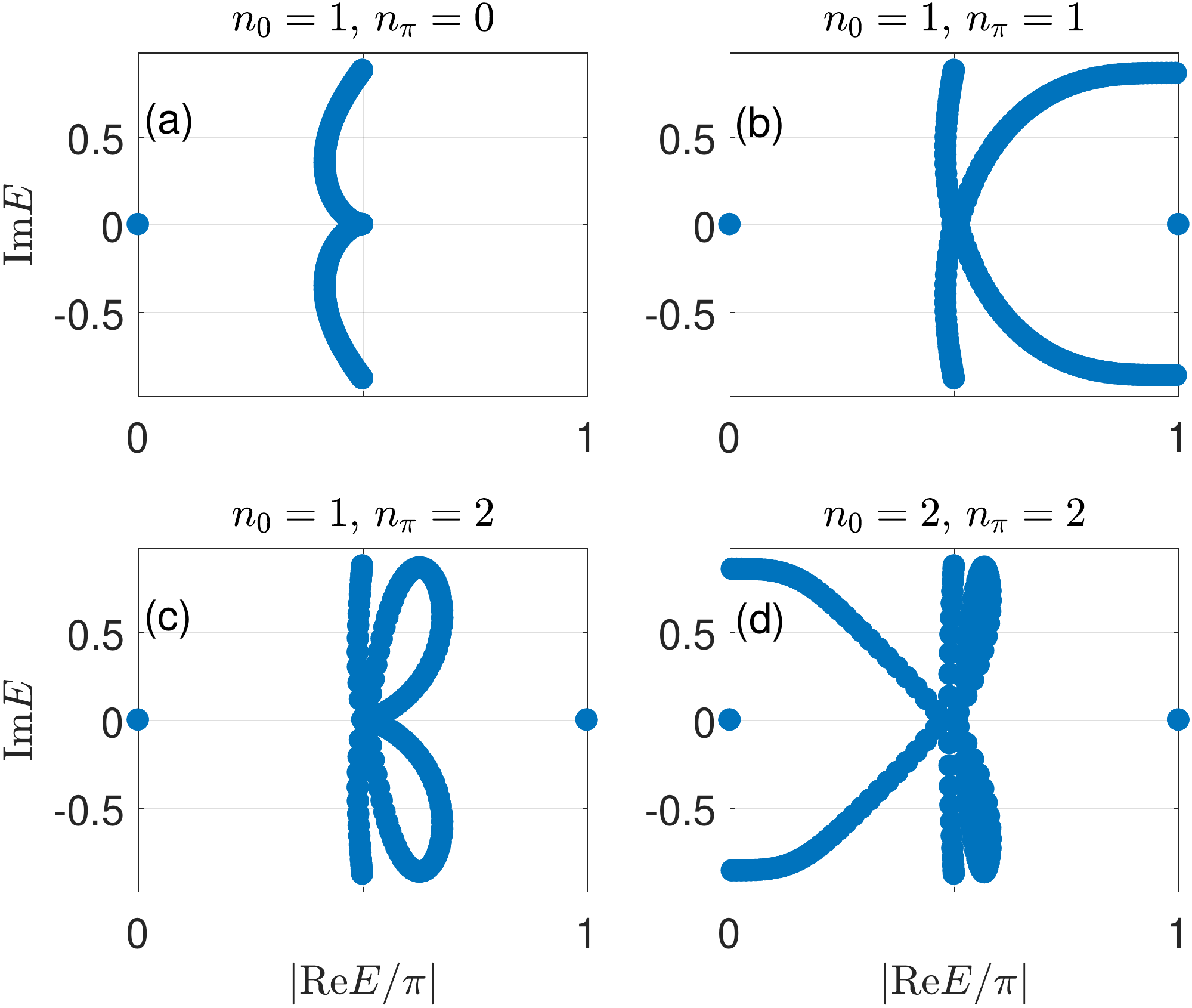}
	\caption{The Floquet spectrum of $\hat{U}$ at separate values of $r_{x}$
		shown in the complex quasienergy plane. The lattice has $N=150$ unit
		cells. Other system parameters are fixed at $\mu=0$, $r_{y}=\frac{\pi}{2}$
		and $\gamma={\rm arccosh}(\sqrt{2})$. Dots at ${\rm Im}E=0$ and
		$|{\rm Re}E|=0$ ($|{\rm Re}E|=\pi$) represent degenerate edge states
		with real quasienergy zero ($\pi$). Other dots denote bulk states
		with complex quasienergies. Panel (a): $r_{x}=\frac{\pi}{2}$, there
		is one pair of edge states at quasienergy $E=0$, and no edge states
		at quasienergy $E=\pi$. Panel (b): $r_{x}=\pi$, there is one
		pair of edge states at quasienergy $E=0$, and one pair of edge states
		at quasienergy $E=\pi$. Panel (c): $r_{x}=\frac{3\pi}{2}$, there
		is one pair of edge states at quasienergy $E=0$, and two pairs of
		edge states at quasienergy $E=\pi$. Panel (d): $r_{x}=2\pi$,
		there are two pairs of edge states at quasienergy $E=0$, and two
		pairs of states at quasienergy $E=\pi$.}
	\label{fig:EdgeState}
\end{figure}

Note in passing that in certain regions of the Floquet spectrum (e.g.
$\pi-\arccos\left[\frac{1}{\cosh(\gamma)}\right]<r_{x}<1+\arccos\left[\frac{1}{\cosh(\gamma)}\right]$
of Fig.~\ref{fig:SpectrumOBC}), it seems that there are no bulk quasienergy gaps at $E=\pm\pi$.
But on the complex quasienergy plane ${\rm Re}E$-${\rm Im}E$, the edge
states at $E=\pm\pi$ are still surrounded by complex
spectrum gaps. To further clarify this point, we show the quasienergy
spectrum at $r_{y}=\frac{\pi}{2}$, $\gamma={\rm arccosh}(\sqrt{2})$
with $r_{x}=\frac{\pi}{2},\pi,\frac{3\pi}{2}$ and $2\pi$ in Fig.~\ref{fig:EdgeState}.
In all these examples, we see that edge states at $E=0$ and $\pm\pi$
indeed have real quasienergies and surrounded by gaps in the complex
quasienergy plane. The survival of these real-quasienergy edge states in the vicinity of
a dissipative bulk may have further applications in achieving robust state manipulations
against environmental effects.

\section{Stroboscopic spin textures}\label{sec:SST}
Besides employing edge states and bulk-edge correspondence, it is
also helpful to find a direct probe to the topological winding numbers
of our PDNHL model. In this section, we suggest to achieve this
by imaging the stroboscopic spin
textures of the system in its dynamics following a sudden quench.
The detections of these spin textures are already available in certain
quantum simulators like ultracold atoms~\cite{TASTExp2018}.
Our approach is based on an earlier proposal regarding the dynamical classification of topological states~\cite{TASTTher2018}. Even though the rigorous proof of the applicability of this proposal to general non-unitary processes has not been fully established, our results strongly indicate that it works at least for our PDNHL model after proper modifications.

\begin{figure}
	\centering
	\includegraphics[width=.47\textwidth]{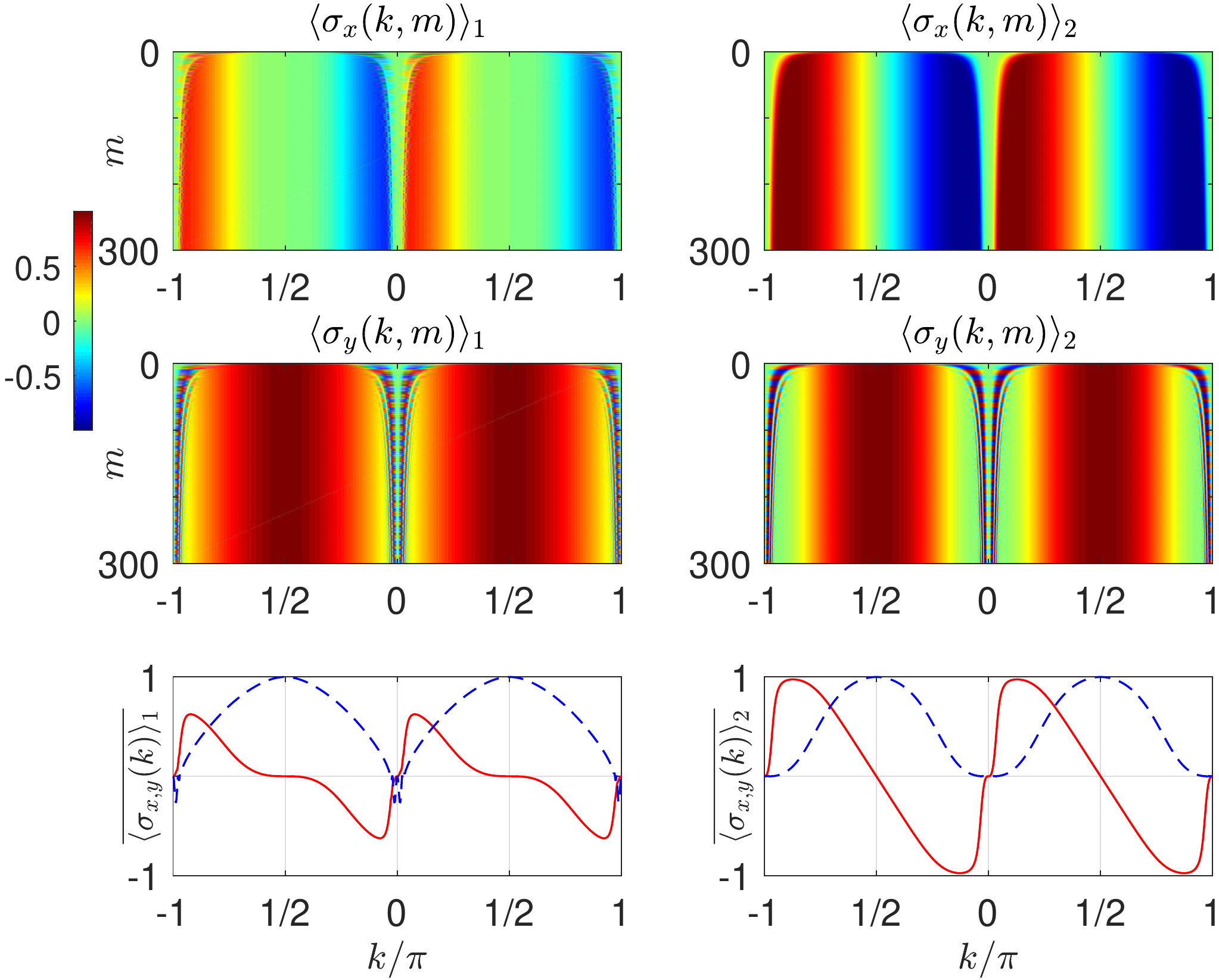}
	\caption{Stroboscopic spin expectation values $\langle\sigma_{j}(k,m)\rangle_{\alpha}$
		(top and middle panels) and stroboscopic time-averaged spin textures
		$\overline{\langle\sigma_{j}(k)\rangle_{\alpha}}$ (bottom panels)
		versus the quasimomentum $k$ for spin components $j=x,y$ in time
		frames $\alpha=1,2$. System parameters are chosen as $\mu=0$, $r_{x}=r_{y}=\frac{\pi}{2}$,
		and $\gamma={\rm arccosh}(\sqrt{2})$. In bottom panels, results for
		$\overline{\langle\sigma_{x}(k)\rangle_{1,2}}$ (red solid lines)
		and $\overline{\langle\sigma_{y}(k)\rangle_{1,2}}$ (blue dashed lines)
		at each $k$ are obtained after averaging over $300$ driving periods.}
	\label{fig:SST1}
\end{figure}

In Ref.~\cite{TASTTher2018}, a dynamical classification of topological
quantum phases was introduced
following a series of three statements: (i) the classification of a generic $d$-dimensional gapped topological phase characterized by integer invariants can be reduced to $(d-1)$-dimensional invariants defined on something called band inversion surfaces (BISs); (ii) the time-averaged spin-polarizations vanish on BISs in the evolution following a quench from a trivial to a topological phase; (iii) the bulk topological property of the post-quench phase is classified by a dynamical topological invariant determined via a dynamical spin-texture field on BISs.

Let's first elaborate a bit more on the meaning of BISs. According to Ref.~\cite{TASTTher2018}, the BISs have a static and a dynamical definition. The static definition comes from the following observations. The minimal model of a $d$-dimensional topological insulator is described by a Bloch Hamiltonian $H({\bf k})=\sum_{i=1}^{d+1}h_i({\bf k})\gamma_i$. Here $\gamma_i$ are Pauli matrices for $d=1,2$ and Dirac $\gamma$ matrices for $d=3,4$. Ref.~\cite{TASTTher2018} suggests to pick out one component, e.g., $h_{d+1}({\bf k})$ from the vector field ${\bf h}({\bf k})=[h_1({\bf k}),...,h_d({\bf k}),h_{d+1}({\bf k})]$, and referring to the remaining components ${\bf h}_{\rm SO}=[h_1({\bf k}),...,h_d({\bf k})]$ as a spin-orbit (SO) field. Then the BISs refer to the $(d-1)$-dimensional surface formed by the collection of all quasimomenta in the first Brillouin zone at which $h_{d+1}({\bf k})=0$. For a $1$-dimensional model, the BISs are formed by discrete $k$-points in the first Brillouin zone, which may also be called ``band inversion points''. On the BISs, a nonzero SO field $h_{\rm SO}({\bf k})$ then serves as a gap-opening perturbation.
Note that according to this definition, the locations and shapes of BISs depend on the choice of $h_{d+1}({\bf k})$, and are therefore not uniquely defined~\cite{TASTTher2018}. As will be discussed later, this nonuniqueness is not an issue, as the topological invariant of the system is solely determined by how a spin texture field (to be defined later) changes across the BISs, not the exact locations and shapes of the BISs themselves. This also provides more room for one to choose $h_{d+1}({\bf k})$ conveniently in experiments without affecting the final detection of topological invariants.

This static definition, however, is \textit{not} the definition of BISs we are going to use. This is because for a non-Hermitian Hamiltonian (or Floquet effective Hamiltonian) $H({\bf k})$, $h_{d+1}({\bf k})$ and the SO field $h_{\rm SO}({\bf k})$ could both be complex-valued, which may not correspond to any physical observables that can be measured experimentally. Furthermore, the choice of $h_{d+1}({\bf k})$ is somewhat ambiguous. This also makes it hard to identify BISs precisely in experiments following this definition.

In the meantime, Ref.~\cite{TASTTher2018} also proposes a dynamical definition of BISs in their Eq.~(4). It states that the BISs are formed by the collection of quasimomenta, at which the dynamical average of spin polarization $\overline{\langle\gamma_i({\bf k})\rangle}=\lim_{\tau\rightarrow\infty}\frac{1}{\tau}\int_0^\tau dt\langle\psi({\bf k},t)|\gamma_i|\psi({\bf k},t)\rangle$ vanishes for every spin component $i=1,2,...,d+1$, where $|\psi({\bf k},t)\rangle$ is the state of the system evolved from a topologically trivial ground state $|\psi({\bf k},0)\rangle$ (e.g., the lowest eigenstate of one spin component $\gamma_i$) prepared at time $t=0$ to a later time $t$.
This definition characterizes BISs by the vanishing of a real-valued spin texture $\overline{\langle\gamma_i({\bf k})\rangle}$, regardless of whether the evolution is unitary or not.

The behaviors of spin textures on the BISs can be further linked to topological properties of the system through the variation of time-averaged spin textures across the BISs. Considering a system prepared initially at a trivial phase, which is described by the ground state of some Hamiltonian $H_0({\bf k})$. At time $t=0$, a sudden quench is applied to the system, such that its dynamical evolution at any time $t>0$ following the quench is governed by a Hamiltonian $H({\bf k})=\sum_{i=1}^{d+1}h_i({\bf k})\gamma_i$ with a topologically nontrivial ground state.

As an example, for a one dimensional system described by a chiral symmetric post-quench Hamiltonian $H(k)=h_{x}(k)\sigma_{x}+h_{y}(k)\sigma_{y}$, the BISs are formed by quasimomenta $k\in[-\pi,\pi)$ at which
\begin{equation}
\overline{\langle\sigma_{j}(k)\rangle}=\lim_{\tau\rightarrow\infty}\frac{1}{\tau}\int_{0}^{\tau}dt\langle\psi(k,t)|\sigma_{j}|\psi(k,t)\rangle=0,\quad j=x,y.
\end{equation}
Here $|\psi(k,t)\rangle$ is the state of the system evolved to time
	$t$ following the quench.
Qualitatively, the averaged spin textures $(\overline{\langle\sigma_x(k)\rangle},\overline{\langle\sigma_y(k)\rangle})$ can vary across each BIS in three possible manners: (i) neither $\overline{\langle\sigma_x(k)\rangle}$ nor $\overline{\langle\sigma_y(k)\rangle}$ changes sign; (ii) both $\overline{\langle\sigma_x(k)\rangle}$ and $\overline{\langle\sigma_y(k)\rangle}$ change signs; (iii) only one component in $\overline{\langle\sigma_x(k)\rangle}$ and $\overline{\langle\sigma_y(k)\rangle}$ changes sign.



To relate the spin textures to the topological property of the post-quench Hamiltonian $H({\bf k})$, Ref.~\cite{TASTTher2018} introduced a dynamical spin texture field ${\boldsymbol g}({\bf k})$, whose components are given by $g_i({\bf k})\equiv-\frac{1}{{\cal N}_{\bf k}}\partial_{k_\bot}\overline{\gamma_i({\bf k})}$, with ${\cal N}_{\bf k}$ being the normalization factor and $k_\bot$ being the quasimomentum perpendicular to the BIS. A topological winding number $w_{d-1}=\sum_j\frac{\Gamma(d/2)}{2\pi^{d/2}}\frac{1}{(d-1)!}\int_{{\rm BIS}_j}{\boldsymbol g}({\bf k})[d{\boldsymbol g}({\bf k})]^{d-1}$ of the dynamical spin texture field ${\boldsymbol g}({\bf k})$ on the BISs was introduced, and proved to be equal to the topological invariant $\nu_d$ characterizing the ground state of the post-quench Hamiltonian $H({\bf k})$~\cite{TASTTher2018}, i.e.,
\begin{equation}
\nu_d=w_{d-1}.\label{eq:TwoInv}
\end{equation}

\begin{figure}
	\centering
	\includegraphics[width=.48\textwidth]{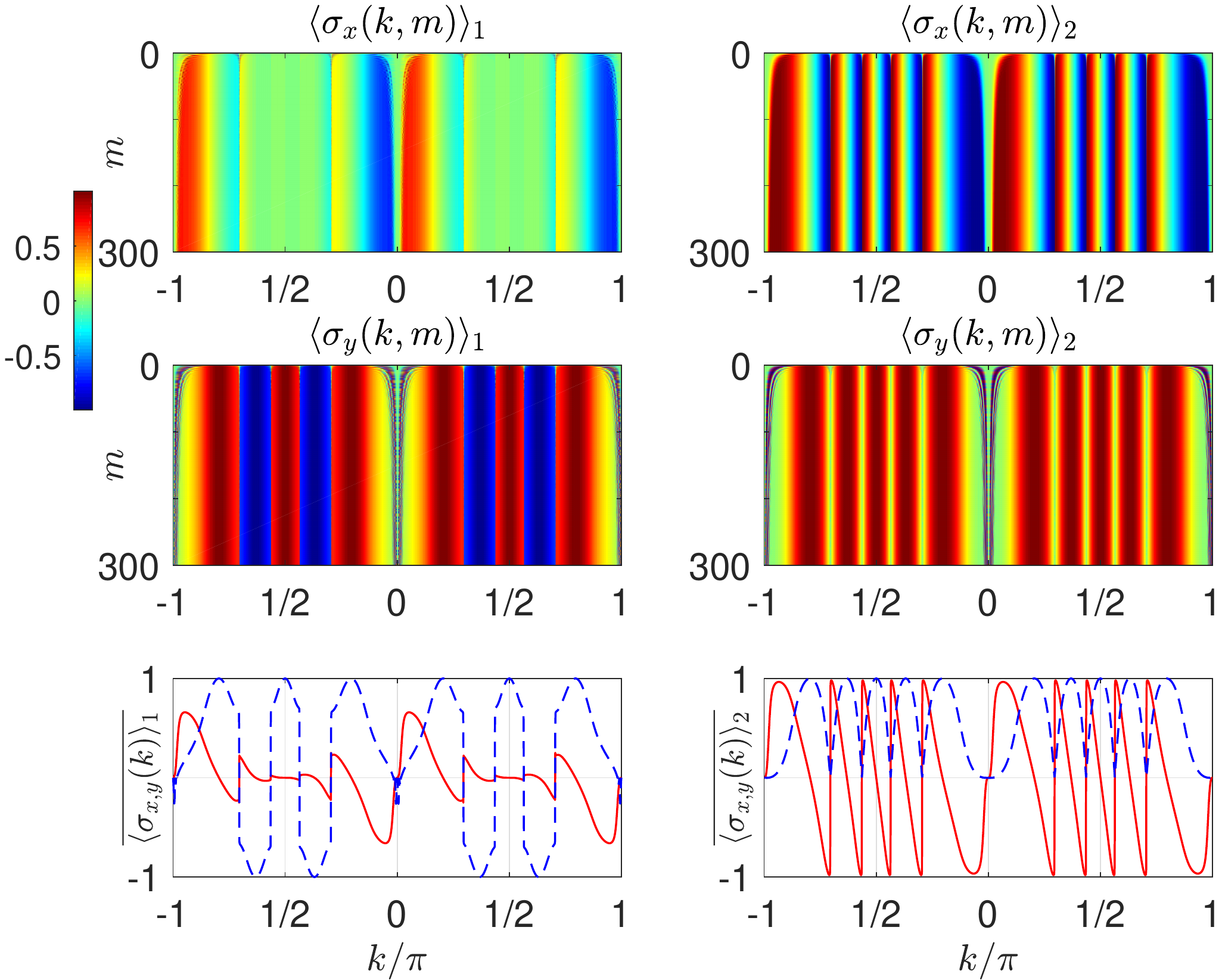}
	\caption{Stroboscopic spin expectation values $\langle\sigma_{j}(k,m)\rangle_{\alpha}$
		(top and middle panels) and stroboscopic time-averaged spin textures
		$\overline{\langle\sigma_{j}(k)\rangle_{\alpha}}$ (bottom panels)
		versus the quasimomentum $k$ for spin components $j=x,y$ in time
		frames $\alpha=1,2$. System parameters are chosen as $\mu=0$, $r_{x}=\frac{5\pi}{2}$,
		$r_{y}=\frac{\pi}{2}$, and $\gamma={\rm arccosh}(\sqrt{2})$. In
		bottom panels, results for $\overline{\langle\sigma_{x}(k)\rangle_{1,2}}$
		(red solid lines) and $\overline{\langle\sigma_{y}(k)\rangle_{1,2}}$
		(blue dashed lines) at each $k$ are obtained after averaging over $300$ driving
		periods.}
	\label{fig:SST2}
\end{figure}

The formalism proposed in Ref.~\cite{TASTTher2018} is applicable to
both closed and open quantum systems. Therefore, it should also be
useful to extract the topological invariants of a non-Hermitian system
in its non-unitary dynamics. To do this, we introduce the stroboscopic
time-averaged spin textures of our PDNHL model as
\begin{equation}
\overline{\langle\sigma_{j}(k)\rangle_{\alpha}}\equiv\lim_{M\rightarrow\infty}\frac{1}{M}\sum_{m=1}^{M}\langle\sigma_{j}(k,m)\rangle_{\alpha},\qquad j=x,y.\label{eq:TAST}
\end{equation}
The stroboscopic spin expectation value $\langle\sigma_{j}(k,m)\rangle_{\alpha}$
is given by
\begin{equation}
\langle\sigma_{j}(k,m)\rangle_{\alpha}=\frac{\langle\phi_{0}|[U_{\alpha}^{\dagger}(k)]^{m}\sigma_{j}U_{\alpha}^{m}(k)|\phi_{0}\rangle}{\langle\phi_{0}|[U_{\alpha}^{\dagger}(k)]^{m}U_{\alpha}^{m}(k)|\phi_{0}\rangle},\quad\alpha=1,2.\label{eq:SSave}
\end{equation}
Here $|\phi_{0}\rangle$ is the initial state, and $U_{\alpha}(k)$
is the Floquet operator in symmetric time frame $\alpha$ as defined
by Eqs.~(\ref{eq:U1k}) and (\ref{eq:U2k}). The normalization factor
$\langle\phi_{0}|[U_{\alpha}^{\dagger}(k)]^{m}U_{\alpha}^{m}(k)|\phi_{0}\rangle$
is introduced since the dynamics is not unitary.
Note that the Pauli matrices here acting on spins instead of sublattice degrees of freedom. In cold atom systems, these degrees of freedom correspond to internal states of atoms, such as the ground hyperfine levels $5 ^2S_{1/2}F = 1$ and $5 ^2S_{1/2}F = 2$ of $^{87}$Rb~\cite{ZhouPRA2018}.

To image the bulk topological invariant of our PDNHL model, we consider dynamics following a sudden quench at the initial time $t=0$ from the prequench Hamiltonian $h_{0}(k)=\sigma_{z}$
to the postquench Hamiltonian $h_{\alpha}(k)=i\ln U_{\alpha}(k)$,
executed separately in the two symmetric time frames $\alpha=1,2$.
The initial states in all cases are chosen to be the ground states
of $h_{0}(k)$ at all $k\in[-\pi,\pi)$, and the spin textures in
postquench dynamics are obtained numerically from Eqs. (\ref{eq:TAST})
and (\ref{eq:SSave}).

Computation examples of $\langle\sigma_{j}(k,m)\rangle_{\alpha}$
and $\overline{\langle\sigma_{j}(k)\rangle_{\alpha}}$ for two hopping
amplitudes $r_{x}=\frac{\pi}{2}$ and $\frac{5\pi}{2}$, with other system parameters
fixed at $\mu=0$, $r_{y}=\frac{\pi}{2}$ and $\gamma={\rm arccosh}(\sqrt{2})$
are presented in Figs.~\ref{fig:SST1} and \ref{fig:SST2}. The results for all $\overline{\langle\sigma_{j}(k)\rangle_{\alpha}}$
are averaged over $M=300$ driving periods.

The dynamical spin texture field of our model in time frame $\alpha$ contains two components $g^{\alpha}_x(k)=-\frac{1}{{\cal N}_k}\partial_{k_\bot}\overline{\langle\sigma_x(k)\rangle_\alpha}$ and $g^{\alpha}_y(k)=-\frac{1}{{\cal N}_k}\partial_{k_\bot}\overline{\langle\sigma_y(k)\rangle_\alpha}$, where the normalization factor ${\cal N}_k=\sqrt{(\partial_{k_\bot}\overline{\langle\sigma_x(k)\rangle_\alpha})^2+(\partial_{k_\bot}\overline{\langle\sigma_y(k)\rangle_\alpha})^2}$. From the numerical results presented in bottom panels of Figs.~\ref{fig:SST1} and \ref{fig:SST2}, we notice that all BISs are extremum points of $\overline{\langle\sigma_y(k)\rangle_\alpha}$. So $g^\alpha_y(k)=0$ on the BISs, and we only need to consider the behavior of $g^\alpha_x(k)$ there, which has the form $[g^\alpha_x(k)]_j=[-{\rm sgn}(\partial_{k_\bot}\overline{\langle\sigma_x(k)\rangle_\alpha})]_j$ at the $j$th BIS. The corresponding topological winding number
\begin{equation}
w^\alpha_0=\sum_j\frac{1}{2}[-{\rm sgn}(\partial_{k_\bot}\overline{\langle\sigma_x(k)\rangle_\alpha})]_j,\label{eq:w0}
\end{equation}
will be quantized if the BISs come in pairs. Then according to the general relation (\ref{eq:TwoInv}), the two winding numbers $w^1_0$ and $w^2_0$ should be related to the topological invariants in the two symmetric time frames $W_1$ and $W_2$~[see Eq.~(\ref{eq:W12})] as:
\begin{equation}
W_1=w^1_0\,,\qquad W_2=w^2_0\,.\label{eq:BSD}
\end{equation}

We can now extract the topological invariants of our PDNHL system from the bottom panels of Figs.~\ref{fig:SST1} and \ref{fig:SST2} with the help of Eqs.~(\ref{eq:w0}) and (\ref{eq:BSD}).

In the bottom panels of Fig.~\ref{fig:SST1}, assuming $k_\bot$ points to left, we have ${\rm sgn}(\partial_{k_\bot}\overline{\langle\sigma_x(k)\rangle_\alpha})=-1$ at the two BISs $k=0,\pi$ in both time frames $\alpha=1,2$. Eqs.~(\ref{eq:w0}) and (\ref{eq:BSD}) then yield $W_1=w^1_0=1$ and $W_2=w^2_0=1$, which are the same as the $W_1$ and $W_2$ calculated by Eq.~(\ref{eq:W12}). Using Eq.~(\ref{eq:W0P}), the two topological invariants $W_0,W_\pi$ of our system are given by $W_0=(w^1_0+w^2_0)/2=1$ and $W_\pi=(w^1_0-w^2_0)/2=0$, which are consistent with the results shown in the phase diagram Fig.~\ref{fig:PhsDiag}.

In the bottom panels of Fig.~\ref{fig:SST2}, we have ${\rm sgn}(\partial_{k_\bot}\overline{\langle\sigma_x(k)\rangle_\alpha})=-1$ at all BISs in both time frames $\alpha=1,2$. Notably, there is one pair of BISs for spin textures in time frame $1$~(bottom left panel of Fig.~\ref{fig:SST2}) and five pairs of BISs for spin textures in time frame $2$~(bottom right panel of Fig.~\ref{fig:SST2}). Eqs.~(\ref{eq:w0}) and (\ref{eq:BSD}) then yield $W_1=w^1_0=1$ and $W_2=w^2_0=5$, which are the same as the $W_1$ and $W_2$ calculated by Eq.~(\ref{eq:W12}). Using Eq.~(\ref{eq:W0P}), the two topological invariants $W_0,W_\pi$ of our system are given by $W_0=(w^1_0+w^2_0)/2=3$ and $W_\pi=(w^1_0-w^2_0)/2=-2$, which are also consistent with the results shown in the phase diagram Fig.~\ref{fig:PhsDiag}.

For system parameters sitting in regular regions of the phase diagram
{[}i.e., $(r_{x},r_{y})\in(0,\pi-\arccos(1/\cosh\gamma))\times(0,\infty)$
and $(r_{x},r_{y})\in(0,\infty)\times(0,\pi)${]}, we have checked
and found the consistency between the winding numbers extracted from
the time-averaged spin textures $\overline{\langle\sigma_{x,y}(k)\rangle_{1,2}}$
and those obtained theoretically from Eq.~(\ref{eq:W12}). Therefore
we conclude that the time-averaged spin textures, proposed
first in Ref.~\cite{TASTTher2018}, can also be a useful tool to image the bulk
invariants of non-Hermitian Floquet topological phases. Experimentally,
the measurement of these spin textures is already available in cold
atom systems~\cite{TASTExp2018}, providing direct signatures of topological invariants
from bulk state dynamics as complementary to the detection of edge
states.

In experiments, the existence of noise and mesoscopic fluctuations may challenge the resolution of spin textures in long-time domains. In Ref.~\cite{TASTTher2018}, the spin textures averaged over a long time have been shown to be robust to slow decay rates. With stronger dissipation effects, our numerical results presented in Figs.~\ref{fig:SST1} and \ref{fig:SST2} suggesting that the spin textures converge quickly to stable patterns within the initial tenth of periods. The topological signatures of our model may then be extracted from the spin textures within a relatively short evolution time, in which the effect of noise may not be fully developed. Indeed, in a recent realization of non-Hermitian photonic quantum walks, spin textures related to the topological properties of the system have also been successfully imaged in a relatively short evolution time~\cite{WangArXiv2018-2}.

More generally, it is interesting to know whether the time averaged
spin textures studied here can also be used to image the topological
invariants of other non-Hermitian systems in different symmetry classes
and at different physical dimensions. These questions are beyond the
scope of this manuscript and we will leave it for future explorations.

\section{Summary and discussion}
In this work, we explored Floquet topological phases in a periodically
driven non-Hermitian lattice model. We found the bulk phase diagram of
the system analytically, with phase boundaries formed by Floquet BTPs
of complex quasienergies versus system parameters.
Each of the phases in the diagram is characterized by a pair of topological
winding numbers, which take integer values and predict the number
of topological edge states at $0$ and $\pi$-quasienergies of the
Floquet spectrum.
Along certain regular directions of the phase diagram,
we also found Floquet topological phases with unlimited winding
numbers and arbitrarily many {\it real}-quasienergy edge states induced solely by the non-Hermiticity of the system.
These edge states are also robust to (non-)Hermitian perturbations which do not break the
chiral symmetry of the system.
We further suggested a dynamical approach to extract the bulk topological winding
numbers of the system by investigating its stroboscopic spin textures~\cite{TASTTher2018}.
A simple connection was observed between the number of times of spin
flips over the Brillouin zone in long time limit and the winding numbers,
suggesting a promising route to detect these non-Hermitian Floquet
topological phases in cold atom systems or other quantum simulators.

Around spectral singularities, noise and mesoscopic fluctuations could have pronounced effects, which may blur the boundary between different topological phases of our model. However, as shown in our phase diagram Fig.~\ref{fig:PhsDiag}, many non-Hermitian Floquet topological phases of our model can be found in wide parameter windows. Deep into each of these parameter window, the topological edge states are well gapped from other bulk states in the corresponding phase. The existence of spectral gaps could then provide protections to the found edge states against noise effects. Experimentally, topological edge states have been observed in non-Hermitian systems~\cite{ZhuArXiv2018}, which also motivate us to ignore the effect of noise in the current study.

By setting $\mu\ne0$ in the model considered in this manuscript,
more asymmetry could appear in its phase diagram, resulting in other
possible non-Hermitian Floquet topological phases. A thorough exploration
of these situations will be left for future study. An extension of
the model studied here to two-dimension may also allow the appearance
of anomalous chiral edge states traversing both the $0$- and $\pi$-quasienergy
gaps of the Floquet spectrum. The topological characterization of
these non-Hermitian anomalous edge states in Floquet systems and their
possible bulk-edge correspondence are also interesting topics for
future explorations.

\section*{Acknowledgement}
J.G. is supported by the Singapore NRF grant No. NRF-NRFI2017-04 (WBS No. R-144-000-378-281) and the Singapore Ministry of Education Academic Research Fund Tier I (WBS No. R-144-000-353-112).

\appendix
\vspace{0.5cm}

\section{Phase boundary diagram: more examples}\label{app:PhsBound}

In this appendix, we give two more examples of the phase boundary
diagrams formed by the trajectories of Floquet BTPs in the $r_{x}$-$r_{y}$
parameter space. The other system parameters are chosen as $\mu=0$,
$\gamma=0.1$ and $\mu=0$, $\gamma=5$ for the two examples presented
here. All the results are obtained from Eq. (\ref{eq:FEP}) in the main
text.

\begin{figure}
	\centering
	\includegraphics[width=.45\textwidth]{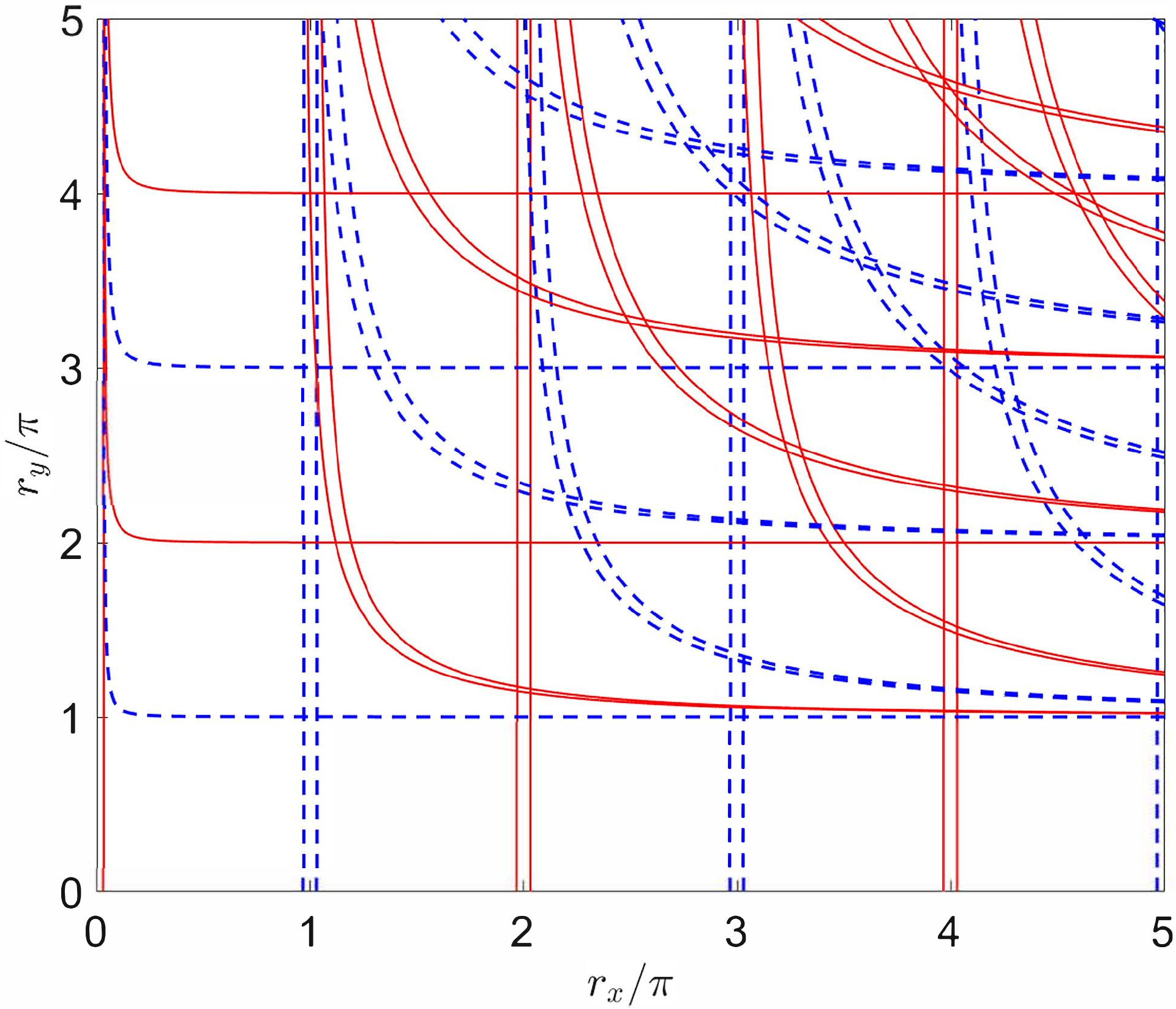}
	\caption{Phase boundary diagram of the PDNHL model in the parameter space $(r_{x},r_{y})\in(0,5\pi]\times(0,5\pi]$,
		with other system parameters fixed at $\mu=0$ and $\gamma=0.1$.
		Trajectories of Floquet BTPs corresponding to spectrum gap closings
		at quasienergy zero ($\pi$) are denoted by red solid (blue dashed)
		lines.}
	\label{fig:PhsBound2}
\end{figure}

In the first example as shown in Fig.~\ref{fig:PhsBound2}, the system is close to its
Hermitian limit, with a small non-Hermitian coupling $\gamma=0.1$.
Compared with the phase diagram of the corresponding Hermitian Floquet
system (see Fig. 1 of Ref.~\cite{ZhouPRA2018}), each phase boundaries now splits
into a pair of closely spaced trajectories formed by Floquet BTPs.
In between, new Floquet topological phases induced by non-Hermiticity
appear, as discussed in the main text. But major parts of the phase
diagram are still dominated by Floquet topological phases carried
over from the system in its Hermitian limit.

\begin{figure}
	\centering
	\includegraphics[width=.45\textwidth]{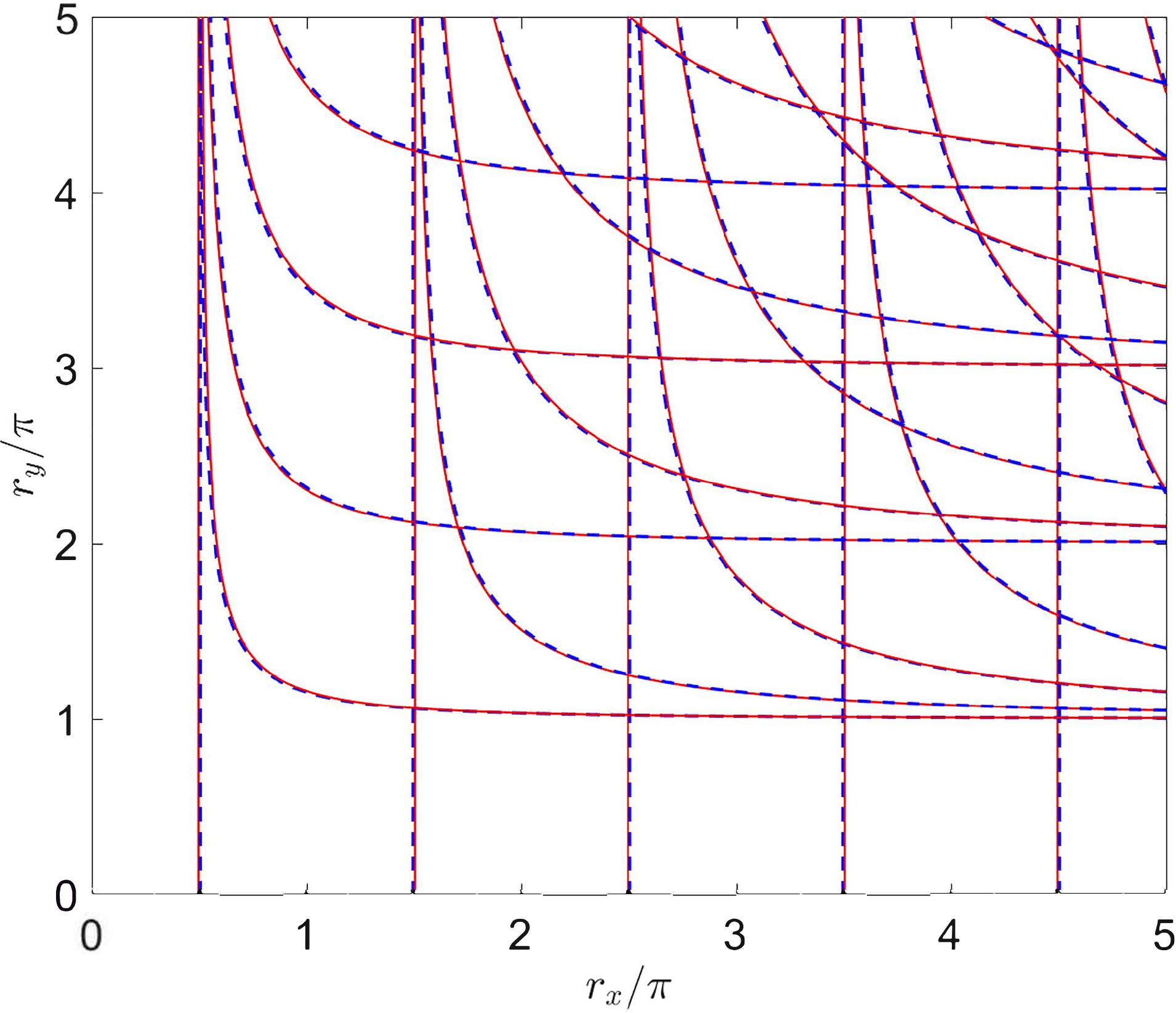}
	\caption{Phase boundary diagram of the PDNHL model in the parameter space $(r_{x},r_{y})\in(0,5\pi]\times(0,5\pi]$,
		with other system parameters fixed at $\mu=0$ and $\gamma=5$. Trajectories
		of Floquet BTPs corresponding to spectrum gap closings at quasienergy
		zero ($\pi$) are denoted by red solid (blue dashed) lines.}
	\label{fig:PhsBound3}
\end{figure}

In the second example as shown in Fig.~\ref{fig:PhsBound3}, the system is close to
a non-Hermitian limit with a relatively large non-Hermitian coupling
$\gamma=5$. New phase boundaries generated by non-Hermitian effects
are now shifted in a way, such that each trajectory of Floquet BTPs
related to a gap closing at quasienergy zero is almost overlapped
with another trajectory of Floquet BTPs related to a gap closing at
quasienergy $\pi$. This means that crossing these phase boundaries
from one side to the other, we will encounter phase transitions accompanied
by spectrum gap closings at both zero and $\pi$ quasienergies simultaneously,
which is very different from the situations of the system in its Hermitian
limit. Therefore, a large non-Hermitian coupling $\gamma$ could introduce
strong modifications to the topological phases of our PDNHL model.

\section{Floquet spectrum and edge states under open boundary conditions: more examples}\label{app:EdgeStat}

In this appendix, we present more examples of the Floquet spectrum
of the PDNHL model under OBC.

\begin{figure}
	\centering
	\includegraphics[width=.45\textwidth]{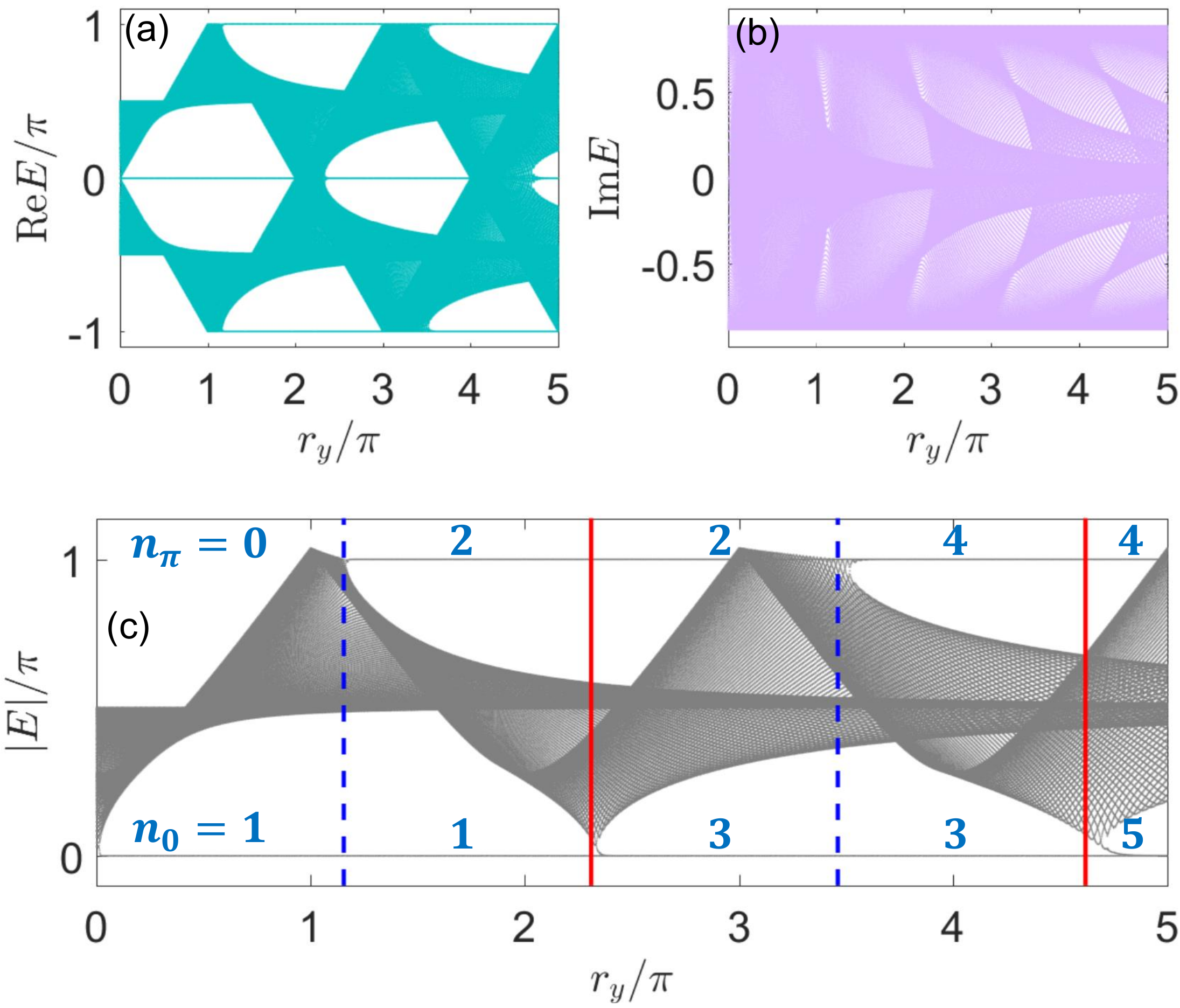}
	\caption{The Floquet spectrum of $\hat{U}$ versus hopping amplitude $r_{y}$
		under OBC. The lattice has $N=300$ unit cells.
		Other system parameters are fixed at $\mu=0$, $r_{x}=\frac{\pi}{2}$
		and $\gamma={\rm arccosh}(\sqrt{2})$. Panels (a) and (b) show the
		real and imaginary parts of the quasienergy $E$. Panel (c) shows
		the absolute values of quasienergy $E$. Red solid (blue dashed) lines
		represent phase boundaries at which the spectrum gaps close at quasienergy
		zero ($\pi$). They are obtained from Eq. (\ref{eq:FEP}) in the main
		text for Floquet BTPs. The integers in light blue in panel (c)
		denote the number of degenerate edge state pairs at quasienergies
		zero ($n_{0}$) and $\pi$ ($n_{\pi}$). Their values are equal to
		the absolute values of winding numbers $W_{0}$ and $W_{\pi}$ as
		shown in Fig.~\ref{fig:WN}(b) of the main text, respectively.}
	\label{fig:SpectrumOBC2}
\end{figure}

\begin{figure}
	\centering
	\includegraphics[width=.45\textwidth]{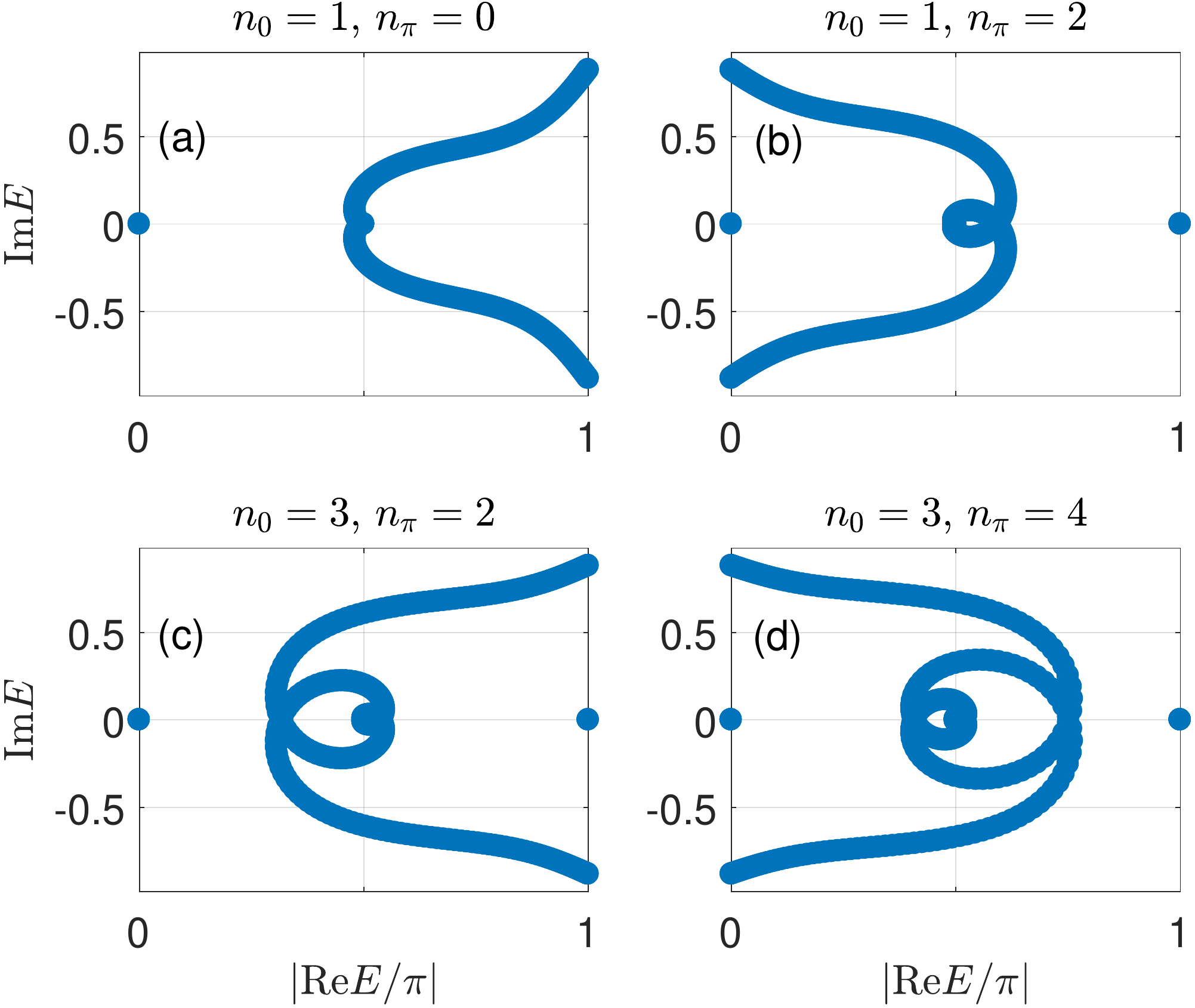}
	\caption{The Floquet spectrum of $\hat{U}$ at separate values of $r_{y}$
		shown in the complex quasienergy plane. The lattice has $N=300$ unit
		cells. Other system parameters are fixed at $\mu=0$, $r_{x}=\frac{\pi}{2}$
		and $\gamma={\rm arccosh}(\sqrt{2})$. Dots at ${\rm Im}E=0$ and
		$|{\rm Re}E|=0$ ($|{\rm Re}E|=\pi$) represent degenerate edge states
		with real quasienergy zero ($\pi$). Other dots denote bulk states
		with complex quasienergies. Panel (a): $r_{y}=\pi$, there is one
		pair of edge states at quasienergy $E=0$, and no edge states at quasienergy
		$E=\pi$. Panel (b): $r_{y}=2\pi$, there is one pair of edge states
		at quasienergy $E=0$, and two pairs of edge states at quasienergy
		$E=\pi$. Panel (c): $r_{y}=3\pi$, there is three pairs of edge
		states at quasienergy $E=0$, and two pairs of edge states at quasienergy
		$E=\pi$. Panel (d): $r_{y}=4\pi$, there are three pairs of edge
		states at quasienergy $E=0$, and four pairs of edge states at quasienergy
		$E=\pi$.}
	\label{fig:EdgeStat2}
\end{figure}
In Fig.~\ref{fig:SpectrumOBC2}, we presented the Floquet spectrum $E$ versus hopping
amplitude $r_{y}$ at fixed values of hopping amplitude $r_{x}=\frac{\pi}{2}$
and non-Hermitian coupling strength $\gamma={\rm arccosh}(\sqrt{2})$
in a lattice of $N=300$ unit cells. In Fig.~\ref{fig:SpectrumOBC2}(c), the red solid and
blue dashed lines correspond to phase boundaries obtained from the
Eq. (\ref{eq:FEP}) for Floquet BTPs. We see that the number of edge
state pairs $(n_{0},n_{\pi})$, as denoted in Fig.~\ref{fig:SpectrumOBC2}(c), changes
across each of the phase boundaries. Furthermore, refer to Fig.~\ref{fig:WN},
we find that $(n_{0},n_{\pi})$ matches exactly the absolute value
of winding number $(|W_{0}|,|W_{\pi}|)$ in each of the corresponding
non-Hermitian Floquet topological phases. Therefore the relation (\ref{eq:BEC})
is verified for the example considered here. In Fig.~\ref{fig:EdgeStat2}, we further
showed the quasienergy spectrum at $r_{x}=\frac{\pi}{2}$, $\gamma={\rm arccosh}(\sqrt{2})$
with $r_{y}=\pi,2\pi,3\pi$ and $4\pi$ in the complex quasienergy
plane. In all these examples, we see that edge states at $E=0$ and
$\pm\pi$ have real quasienergies and are surrounded by gaps in the
complex quasienergy plane.

\newpage{}

\end{document}